\documentclass{amsart}
\usepackage{amsmath}%
\usepackage{amsfonts}%
\usepackage{amssymb}%
\usepackage{graphicx}
\usepackage[authoryear]{natbib}
\def\MMMalphabar{0.005860}
\def\MMMalphabarSE{0.000613}
\def\MMMeta{0.049496}
\def\MMMetaSE{0.002968}
\def\MMMloglkhd{1028.776695}
\def\BStheta{0.130386814}
\def\BSthetaSE{0.003405}
\def\BSloglkhd{1019.842904}
\newtheorem{theorem}{Theorem}
\theoremstyle{plain}

\newtheorem{ass}{Assumption}

\newtheorem{corollary}{Corollary}

\newtheorem{definition}{Definition}

\newtheorem{lemma}{Lemma}

\newtheorem*{refproof*}{Proof}

\begin{document}
\title[Less-Expensive Valuation of Long Term Annuities]{Less-Expensive Valuation of Long Term Annuities Linked to Mortality, Cash and Equity}
\author{Kevin Fergusson}
\curraddr[Kevin Fergusson]
{University of Melbourne \newline%
\indent Victoria 3010, Australia}%
\email[Kevin Fergusson]{kevin.fergusson@unimelb.edu.au}%
\urladdr{http://fbe.unimelb.edu.au/economics/ACT}
\author{Eckhard Platen}
\curraddr[Eckhard Platen]{ University of Technology, Sydney
\newline%
\indent PO Box 123, Broadway NSW 2007, Australia}%
\email[Eckhard Platen]{eckhard.platen@uts.edu.au}%
\urladdr{http://www.business.uts.edu.au}
\date{\today{}}
\thanks{This research is supported by an Australian Government Research Training Program Scholarship.}
\begin{abstract}
This paper proposes a paradigm shift in the valuation of long term annuities, away from classical no-arbitrage valuation towards
valuation under the real world probability measure. Furthermore, we apply this valuation method to two examples of annuity products,
one having annual payments linked to a mortality index and the savings account and the other having annual payments linked to a mortality index and an equity index with a guarantee that is linked to the same mortality index and the savings account. Out-of-sample hedge simulations demonstrate the effectiveness of real world valuation.

In contrast to
risk neutral valuation, which is a form of relative valuation, the long term average excess return of the equity
market comes into play. Instead of the savings account, the num\'eraire portfolio is employed as the
fundamental unit of value in the analysis. The
num\'eraire portfolio is the strictly positive, tradable portfolio
that when used as benchmark makes all benchmarked nonnegative
portfolios supermartingales. The
benchmarked real world value of a benchmarked contingent claim
equals its real world conditional expectation. This yields the minimal
possible value for its hedgeable part and minimizes the fluctuations for its
benchmarked hedge error. Under classical assumptions, actuarial and risk neutral valuation emerge
as special cases of the proposed real world valuation. In long term
liability and asset valuation, the proposed real world valuation can lead to
significantly lower values than suggested by classical
approaches when an equivalent risk neutral probability measure does not exist.
\end{abstract}

\maketitle

\section{Introduction}
Long dated contingent claims are relevant in insurance, pension
fund management and derivative valuation. This paper proposes a
paradigm shift in the valuation of long term contracts, away from
classical no-arbitrage valuation, towards valuation under the real
world probability measure. In contrast to risk neutral valuation, which is a form of relative valuation,
the long term average trend of the equity market above the fixed income money market, known as the
equity premium, is coming into play in the proposed real world valuation.
A benchmark, the num\'eraire portfolio, is employed as the
fundamental unit of value in the analysis, replacing the
savings account. The num\'eraire
portfolio is the strictly positive, tradable portfolio that when
used as benchmark makes all benchmarked nonnegative portfolios
supermartingales. This means, their current benchmarked values are
greater than, or equal to, their expected future benchmarked values.
Furthermore, the benchmarked real world value of a benchmarked contingent
claim is proposed to equal its real world conditional expectation.
This yields the minimal possible value for its hedgeable part of the benchmarked contingent claim and
minimizes the fluctuation of its benchmarked hedge error. It turns out
that the pooled total benchmarked hedge error of a well diversified book of contracts issued by an
insurance company can practically vanish due to
diversification when the number of contracts becomes large.
Classical actuarial and risk neutral valuation emerge as special
cases of the proposed real world
valuation methodology when classical modeling assumptions are imposed. In long term asset and liability
valuation, real world valuation can lead to significantly lower
values than suggested by classical valuation arguments when the existence of some equivalent risk neutral
probability measure is not requested. A wider and more
realistic modeling framework then becomes available which allows this phenomenon to be exploited.

The {\em benchmark approach},
described in
\cite{PlHe2010}, proposes such a framework. Instead of relying
on the domestic savings account as the reference unit, a benchmark in form of the best
performing, tradable strictly positive portfolio is chosen as num\'eraire. More precisely, it is proposed to employ the {\em
num\'eraire portfolio} as benchmark, whose origin can be traced back to
\cite{Long90}, and which is equal to, in general, the growth optimal portfolio; see \cite{Kelly56}.

In recent years the problem of accurately valuing long term
assets and liabilities, held by insurance companies, banks and
pension funds, has become increasingly important. How these
institutions perform such valuations often remains unclear.
However, the recent experience with low interest rate environments suggests that some major changes are due in these industries.
One possible explanation, to be explored in this article,
is that the risk neutral valuation paradigm itself may be inherently
flawed, especially when it is applied to the valuation of long
term contracts. It leads to a more expensive production method than necessary, as will be explained in the current paper.

The structure of the paper is as follows: Section~\ref{sec:val2}
gives a brief survey on the literature about valuation methods in
insurance and finance. Section~\ref{sec:val4}
introduces the benchmark approach. Real world valuation is described
in Section~\ref{sec:val5}. Two examples on real world valuation and hedging of long term annuities, with annual payments linked to a mortality index and either a savings account or an equity index, are illustrated in Section~\ref{sec:val8}. Section~\ref{sec:val10} concludes.

\section{Valuation Methods for Long Term Contracts}
\label{sec:val2}
One of the most dynamic areas in the current risk management
literature is the valuation of long term contracts, including variable annuities. The latter represent
long term contracts with payoffs that depend on insured events and
on underlying assets that are traded in financial markets. The valuation methods can be categorised into three main types: actuarial valuation or expected present value, risk neutral valuation and utility maximization valuation.

{\noindent \em Actuarial Valuation (Expected Present Value)} 

One of the pioneers of calculating present values of contingent claims for life insurance companies was James Dodson, whose work is described in the historical accounts of \cite{Camp2016} and \cite{Dods1756}. The application of such methods to with-profits policies ensued. In the late 1960s US life insurers entered the variable annuity market, as mentioned in \cite{Sloane70}, where such products required assumptions on the long term behaviour of equity markets. Many authors have analysed various actuarial models of the long term
evolution of stochastic equity markets, such as \cite{Wise84} and \cite{Wilkie85,Wilkie87,Wilkie95}.

Since the work of \cite{Redi1952}, the matching of well-defined cash flows with liquidly traded ones,
while minimizing the risk of reserves, has been a widely used
valuation method in insurance. For instance, \cite{Wise84b,Wise84,Wise87a,Wise87b,Wise89}, \cite{Wilkie85}
and \cite{KeelMu95} study contracts when a perfect match is not
possible.

{\noindent \em Risk Neutral Valuation} 

The main stream of research, however, follows the concept of
no-arbitrage valuation in the sense of \cite{Ross76} and
\cite{HarrisonKr79}. This approach has been widely used in
finance, where it appears in the guise of risk neutral valuation.
The earliest applications of no-arbitrage valuation to variable
annuities are in the papers by \cite{BrennanSc76,BrennanSc79} and \cite{BoyleSc77}, which
 extend the Black-Scholes-Merton option valuation (see
\cite{BlackSc73} and \cite{Merton73b}) to the case of
equity-linked insurance contracts. 

The Fundamental Theorem of Asset Pricing, in its most general form formulated
by \cite{DelbaenSc98}, establishes a correspondence between the
``no free lunch with vanishing risk" no-arbitrage concept and the
existence of an equivalent risk neutral probability measure. This
important result demonstrates that theoretically founded classical no-arbitrage pricing
requires the restrictive assumption that an equivalent risk
neutral probability measure must exist. In such a setting,
the effect of
stochastic interest rates on the risk neutral value of a guarantee
has been discussed by many authors, such as \cite{BacinelloOr93,BacinelloOr96}, \cite{AasePe94} and \cite{HuangCa04,HuangCa05}.

{\noindent \em Risk Neutral Valuation for Incomplete Markets}

In reality, one has to deal with the fact that markets are
incomplete and insurance payments are not fully hedgeable. The choice of a risk neutral pricing measure is,
therefore, not unique, as pointed out by \cite{FollmerSo86} and
\cite{FollmerSc91}, for example.
\cite{HofmannPlSc92}, \cite{GerberSh94},
\cite{Gerber97} and \cite{JaimungalYo05}. \cite{DuffieRi91} and
\cite{Schweizer92} address this issue by suggesting certain
mean-variance hedging methods based on a form of variance- or
risk-minimizing objective, assuming the existence of a particular
risk neutral measure. In the latter case, the so-called minimal
equivalent martingale measure, due to \cite{FollmerSc91}, emerges
as the pricing measure. This valuation method is also known as local risk
minimization and was considered by \cite{Moller98,Moller01}, \cite{Schweizer01} and
\cite{DahlMo06} for the valuation of insurance products. 

{\noindent \em Expected Utility Maximization}

Another approach involves the maximization of expected terminal
utility, see \cite{KaratzasShLeXu91}, \cite{KramkovSc99}
and \cite{Delbaen_6_02}.
In this case the valuation is based on a
particular form of utility indifference pricing. This form of
valuation has been applied by \cite{HodgesNe89} and later by
\cite{Davis97}. It has been used to value equity-linked insurance
products by \cite{YoungZa02d,YoungZa02i},
\cite{Young03} and \cite{MooreYo03}.

Typically in the context of some expected utility maximization there is an ongoing debate on the links between the valuation of
insurance liabilities and financial economics for which the
reader can be refered to \cite{Reitano97}, \cite{Longley98},
\cite{BabbelMe98}, \cite{Moller98,Moller02}, \cite{PhillipsCuAl98},
\cite{Girard00}, \cite{Lane00}
and \cite{Wang00,Wang02}. Equilibrium modeling from a macro-economic
perspective has been the focus of a line of research
that can be traced back to \cite{Debreu82}, \cite{Starr97} and \cite{Duffie92}.

{\noindent \em Stochastic Mortality Rates}

Note that stochastic mortality rates are easily incorporated in
the pricing of insurance products as demonstrated by
\cite{MilevskyPr01}, \cite{Dahl04}, \cite{KirchMe05}, \cite{CairnsBlDo06a,CairnsBlDo06p,CairnsBlDo08},
\cite{Biffis05}, \cite{MelnikovRo06,MelnikovRo08} and \cite{JalenMa08}. Most
of these authors assume that the market is complete with respect
to mortality risk, which means that it can be removed by
diversification.

{\noindent \em Stochastic Discount Factors}

Several no-arbitrage pricing concepts have been popular in
finance that are equivalent to the risk neutral approach. For instance, \cite{Cochrane01} employs the
notion of a stochastic discount factor. The use of a state-price
density, a deflator or a pricing kernel have been considered
by \cite{Constantinides92}, \cite{Cochrane01} and
\cite{Duffie92}, respectively. Another way of describing classical no-arbitrage pricing was
pioneered by \cite{Long90} and further developed in
\cite{BajeuxPo97} and \cite{Becherer01}, who use the num\'eraire
portfolio as num\'eraire instead of the savings account, and
employ the real world probability measure as pricing measure to recover risk neutral prices.

{\noindent \em Real World Pricing under the Benchmark Approach}

The previous line of research involving the num\'eraire portfolio comes closest to the form of real world valuation
proposed under the benchmark approach in \cite{Plat2002b} and \cite{PlHe2010}. The primary difference is that the benchmark approach
does no longer assume the existence of an equivalent risk neutral probability
measure. In so doing it allows for a much richer class of models to
be available for consideration and permits several self-financing portfolios
to replicate the same contingent claim, where it can select the least expensive one as corresponding value process.

The benchmark approach employs the
best performing, strictly positive, tradable portfolio as
benchmark and makes it the central reference unit for modeling, pricing and hedging.

All valuations are performed under the real world
probability measure and, therefore, labelled
``real world pricing".
In a complete market where an equivalent risk neutral probability measure exists real world pricing yields the same price as risk neutral pricing. When there is no equivalent risk neutral probability measure in the market model, then risk neutral prices can still be employed without generating any economically meaningful arbitrage but these may be more expensive than the respective real world prices.

In~\cite{DuPl16} the concept of benchmarked risk minimization has been introduced, which yields via the real world price the minimal value for the hedgeable part of a not fully hedgeable contingent claim and minimizes the fluctuations of the profit and losses when denominated in units of the num\'eraire portfolio.
Risk minimization that is close to the previously mentioned concept of local risk minimization of \cite{FollmerSc91} and \cite{FollmerSo86} was studied under the benchmark approach in \cite{BiCrPl2014}.

\section{Benchmark Approach}

   \label{sec:val4}
Within this and the following section we give a brief survey about the benchmark approach, which goes beyond results presented in \cite{PlHe2010} and underpins our findings. Consider a market comprising a finite number $J+1$ of
primary security accounts. An example of such a security could
be an account containing shares of a company with all dividends
reinvested in that stock. A savings account held in some currency
is another example of a primary security account. In reality,
time is continuous and this paper considers continuous time models. These can provide compact and elegant
mathematical descriptions of asset value dynamics. We work on a filtered probability space $(\Omega , {\mathcal{F}},{\underline{\mathcal{F}}} , P )$ with filtration ${\underline{\mathcal{F}}} = ({\mathcal{F}}_t)_{t\ge 0}$ satisfying the usual conditions, as in \cite{KaratzasSh91}.

This section introduces the benchmark approach with its concept of real world pricing.
The key assumption is that there exists a best performing, strictly positive, tradable
portfolio in the given investment universe, which we specify later on as the num\'eraire portfolio. This benchmark
portfolio can be interpreted as a universal currency. Its existence turns out to be sufficient
for the formulation of powerful results concerning diversification,
portfolio optimization and valuation.

The {\em benchmarked value\/} of a security represents its
value denominated in units of the benchmark portfolio. Denote
by ${\hat{S}}^j_t$ the benchmarked value of the $j$th {\em primary
security account}, $j \in \{ 0,1,\ldots ,J\}$, at time $t \geq 0$.
The $0$-th primary security account is chosen to be the savings account of the domestic currency. The particular dynamics of the primary security accounts are not
important for the formulation of several statements presented below. For simplicity,
taxes and transaction costs are neglected in the paper.

The market participants can form self-financing portfolios with
primary security accounts as constituents. A portfolio at time
$t$ is characterized by the number $\delta^j_t$ of units held in
the $j$th primary security account, $j \in \{ 0,1,2,\ldots ,J \}$, $t \geq 0$.
Assume for any given {\em strategy \/} $\delta=\{\delta_t=(\delta^0_t,
\delta^1_t,\ldots , \delta^J_t )^\top, t \geq 0 \}$ that the values
$\delta^0_t,\delta^1_t,\ldots ,\delta^J_t $ depend only on
information available at the time $t$. The
value of the benchmarked portfolio, which means its value
denominated in units of the benchmark, is given at time $t$ by the sum
\begin{equation} \label{dt2.5'}
 {\hat{S}}^\delta_t = \sum_{j=0}^J \delta^j_t\,{\hat{S}}^j_t,
 \end{equation}
for $t \geq 0$. Since there is only finite total wealth available in
the market, the paper considers only strategies whose associated benchmarked
portfolio values remain finite at all times.

Let $E_t(X)=E(X|{\mathcal{F}}_t)$ denote the expectation of a random variable $X$
under the real world probability measure $P$, conditioned on the
information available at time $t$ captured by ${\mathcal{F}}_t$ (for example, see Section 8 of Chapter 1 of \cite{Shiryaev84}). This allows us to formulate
the main assumption of the benchmark approach as follows:

\begin{ass}
\label{ass:dt3.2b}
There exists a strictly positive benchmark portfolio,
called the num\-\'eraire portfolio,
such that each benchmarked nonnegative portfolio ${\hat{S}}^\delta_t$ forms a
supermartingale, which means that
 \begin{equation} 
 \label{bval2}
 {\hat{S}}^\delta_t \geq E_t({\hat{S}}^\delta_s)
 \end{equation} 
for all $0 \leq t \leq s < \infty$.
\end{ass}

Inequality \eqref{bval2} can be referred to as the {\em supermartingale
property\/} of benchmarked securities. It is obvious that the benchmark represents in the sense of Inequality \eqref{bval2}
the best performing portfolio,
forcing all benchmarked nonnegative portfolios in the mean
downward or having no trend. In general the assumed num\'eraire portfolio
coincides with the growth optimal portfolio, which is
the portfolio that maximizes expected logarithmic utility, see \cite{Kelly56}. Since
only the existence of the num\'eraire portfolio is requested,
the benchmark approach reaches
beyond the classical no-arbitrage modeling world.

According to Assumption~\ref{ass:dt3.2b} the current benchmarked value of a nonnegative
portfolio is greater than or equal to its expected future
benchmarked values. 
Assumption~\ref{ass:dt3.2b} guarantees
several essential properties of a financial
market model without assuming a particular dynamics for the asset values.
For example, it implies the absence of economically
meaningful arbitrage by ensuring that any strictly positive portfolio remains finite at any finite time because the best performing portfolio has this property.
As a consequence of the supermartingale property \eqref{bval2}
and because a nonnegative supermartingale that reaches zero is absorbed at zero, no wealth
can be created from zero initial capital under limited liability.

For the classical risk neutral valuation, the corresponding no-arbitrage concept is formalised as ``no free lunch with vanishing risk" (NFLVR), see \cite{DeSc1994b}. The benchmark approach assumes that the portfolio cannot explode, which is equivalent to the ``no unbounded profits with bounded risk" (NUPBR) concept, see \cite{KaratzasKa07}. An equivalent martingale measure is not required to exist and, therefore, benchmarked portfolio strategies are permitted to form strict supermartingales, a phenomenon which we exploit in the current paper.

Another fundamental property that follows directly from the supermartingale property \eqref{bval2} is that the benchmark portfolio is
unique. To see this, consider two strictly positive portfolios
that are supposed to represent the benchmark. The first
portfolio, when expressed in units of the second one, must satisfy
the supermartingale property \eqref{bval2}. By the same argument,
the second portfolio, when expressed in units of the first one,
must also satisfy the supermartingale property. Consequently, by
Jensen's inequality both portfolios must be identical. Thus, the
value process of the benchmark that starts with given
strictly positive initial capital is unique. Due to possible redundancies in
the set of primary security accounts, this does not imply
uniqueness for the trading strategy generating the benchmark
portfolio.

Assumption~\ref{ass:dt3.2b} is satisfied for most reasonable
financial market models. It simply asserts the existence of a best
performing portfolio that does not ``explode". This requirement
can be interpreted as the absence of economically meaningful
arbitrage. In Theorem 14.1.7 of Chapter 14 of \cite{PlHe2010}, Assumption~\ref{ass:dt3.2b} has been verified for jump diffusion markets,
which cover a wide range of possible market
dynamics. 
\cite{KaratzasKa07} show that Assumption~\ref{ass:dt3.2b} is satisfied for any reasonable semimartingale model.
Note that Assumption~\ref{ass:dt3.2b} permits us to model
benchmarked primary security accounts that are not martingales.
This is necessary for realistic long term market modeling,
as will be demonstrated in Section~\ref{sec:val8}.

By referring to results in \cite{Platen05dp}, \cite{LePl06a}, \cite{PlatenRe12} and \cite{PlRe17},
one can say that the benchmark portfolio is not only a
theoretical construct, but can be approximated by well diversified
portfolios, e.g. by the MSCI world stock index for the global equity market or the S$\&$P500 total return index for the
US equity market.

A special type of security emerges when equality holds in
relation \eqref{bval2}.

\begin{definition}
 \label{def:bval4} \quad
A security is called {\em fair\/} if its benchmarked value
${\hat{V}}_t$ forms a martingale, that is, the current value of the process $\hat{V}$
is the best forecast of its future values, which means that,
 \begin{equation} 
 \label{bval6}
 {\hat{V}}_t = E_t\left( {\hat{V}}_s\right)
 \end{equation} 
for all $0 \leq t \leq s < \infty$.
\end{definition}

Note that the above notion of a fair
security is employing the best performing
portfolio, the benchmark. 
The benchmark approach allows us to consider
securities that are not fair. This important flexibility is missing in
the classical no-arbitrage approach and
will be required when modeling the
market realistically over long time periods.

\section{Real World Pricing}
  \label{sec:val5}
 
As stated earlier, the most obvious difference between the benchmark
approach and the classical risk neutral approach is the choice of
pricing measure. The former uses the real world probability measure with the
num\'eraire portfolio as reference unit
for valuation, while the savings account is the chosen num\'eraire under the risk
neutral approach, which assumes the existence of an equivalent risk neutral probability measure. The assumption is additionally imposed to our Assumption~\ref{ass:dt3.2b} and, therefore, reduces significantly the class of models and phenomena considered. The supermartingale property \eqref{bval2}
ensures that the expected return of a benchmarked nonnegative
portfolio can be at most zero. In the case of a fair benchmarked
portfolio, the expected return is precisely zero. The current
benchmarked value of such a portfolio is, therefore, the
best forecast of its benchmarked future values. The risk neutral approach assumes that the savings account is fair, which seems to be at odds with evidence; see for example \cite{BaGrPl2015} and \cite{BaIgPl2016}.

Under the benchmark approach, there can be many supermartingales that approach the same future random value.
Within a family of
nonnegative supermartingales, the supermartingale with the smallest initial value turns out to be the
corresponding martingale; see Proposition 3.3 in~\cite{DuPl16}. This basic fact allows us to deduce
directly the following {\em Law of the Minimal Price}:

\begin{theorem}
 \label{the:dt4'.4} \quad {\em (Law of the Minimal Price)} \quad
If a fair portfolio replicates a given nonnegative payoff at some future time,
then this portfolio represents
the minimal replicating portfolio among all nonnegative portfolios
that replicate this payoff.
 \end{theorem}

For a given payoff there may exist self-financing hedge portfolios that are not fair.
Consequently, the classical Law of One Price (see, for example, \cite{Tayl02})
does no longer hold under the benchmark approach. However, the above Law of the Minimal
Price provides instead a consistent, unique value system for all
hedgeable contracts with finite expected benchmarked payoffs.

It follows for a given hedgeable payoff that the corresponding fair hedge portfolio
represents the least expensive hedge portfolio. From an economic
point of view investors prefer more to less and this is, therefore, also the correct value in a liquid, competitive
market. As will be demonstrated in Section~\ref{sec:val8}, there may exist several
self-financing portfolios that hedge one and the same payoff. It is
the fair portfolio that hedges the payoff at minimal cost. 
We emphasize that risk neutral valuation
based purely on hedging via classical no-arbitrage arguments, see
\cite{Ross76} and \cite{HarrisonKr79}, may lead to more expensive values than those given
by the corresponding fair value.

Now, consider the problem of valuing a given payoff to be delivered
at a maturity date $T \in (0,\infty)$. Define
a benchmarked {\em contingent claim\/} ${\hat{H}}_T$ as a nonnegative
payoff denominated in units of the benchmark portfolio with finite expectation
 \begin{equation} 
 \label{dt4'.a}
 E_0 \left( {\hat{H}}_T\right)  < \infty.
 \end{equation}

If for a benchmarked contingent claim ${\hat{H}}_T$, $T \in
(0, \infty)$, there exists a benchmarked fair portfolio ${\hat{S}}^{\delta_{\hat{H}_T}}$,
which replicates this claim at maturity $T$, that is ${\hat{H}}_T =
{\hat{S}}^{\delta_{\hat{H}_T}}_T$, then, by the above Law of the Minimal Price, its
minimal replicating value process is at time $t \in
[0,T]$ given by the real world conditional expectation
 \begin{equation} 
 \label{dt4'.63}
 {\hat{S}}^{\delta_{\hat{H}_T}}_t = E_t
 \left( {\hat{H}}_T \right) .
 \end{equation}
Multiplying both sides of equation \eqref{dt4'.63} by the
value of the benchmark portfolio in domestic currency at time $t$, denoted by $S^*_t$,
one obtains the {\em real
world valuation formula\/}
 \begin{equation} 
 \label{val3.3}
 S^{\delta_{\hat{H}_T}}_t = \hat{S}^{\delta_{\hat{H}_T}}_t \, S^*_t = S^*_t\,
 E_t \left( \frac{H_T}{S^*_T} \right) ,
 \end{equation}
where $H_T={\hat{H}}_T\,S^*_T$ is the payoff denominated in domestic
currency and $S^{\delta_{\hat{H}_T}}_t$ the fair value at time $t \in [0,T]$ denominated in domestic currency.
Note that the benchmark portfolio can be obtained by the product
$S^*_t=(\hat{S}^0_t)^{-1}\, S^0_t$ of the inverse of the benchmarked savings account $\hat{S}^0_t$
and the value $S^0_t$ of this savings account denominated in
domestic currency.

Formula \eqref{val3.3} is called the real world valuation formula
because it involves the conditional expectation $E_t$ with
respect to the real world probability measure $P$. It only requires
the existence of the num\'eraire portfolio and the finiteness of
the expectation in \eqref{dt4'.a}. These two conditions can
hardly be weakened. By introducing the concept of benchmarked risk minimization in~\cite{DuPl16} it has been shown that the above real
world valuation formula also provides the natural valuation for
nonhedgeable contingent claims when one aims to diversify as much as possible nonhedgeable parts of contingent claims.

An important application for the real world pricing formula
\eqref{val3.3} arises when $H_T$ is independent of $S^*_T$, which leads to
the {\em actuarial valuation formula\/}
  \begin{equation} 
 \label{dt4'.63a}
S^{\delta_{{\hat{H}}_T}}_t =P(t,T)\,E_t(H_T).
 \end{equation} 
The derivation of \eqref{dt4'.63a} from \eqref{val3.3}
exploits the simple fact that the expectation of a product of independent random
variables equals the product of their expectations.
One discounts in \eqref{dt4'.63a} by multiplying the real world expectation $E_t(H_T)$ with the fair zero coupon bond value
  \begin{equation} 
 \label{dt4'.63b}
 P(t,T) = S^*_t\,E_t\left( (S^*_T)^{-1}\right) .
 \end{equation} 
The actuarial valuation formula \eqref{dt4'.63a} has been used as a valuation rule by actuaries for centuries to
determine the net present value of a claim. This important formula follows here as a direct
consequence of real world pricing, confirming actuarial intuition and experience.

The following discussion aims to highlight the link between real world valuation and risk
neutral valuation. Risk neutral valuation uses as its num\'eraire the domestic savings
account process $S^0=\{S^0_t \, , \, t \geq 0 \}$, denominated in units of
the domestic currency. Under certain assumptions,
which will be described below, one can derive risk neutral values
from real world values by rewriting the real world valuation formula
\eqref{val3.3} in the form
 \begin{equation} 
 \label{dt4'.b7}
 S^{\delta_{{\hat{H}}_T}}_t = E_t \left( \frac{\Lambda_T}{\Lambda_t}
 \,\frac{S^0_t}{S^0_T} \,H_T \right) ,
 \end{equation} 
employing the benchmarked, normalized savings account
$\Lambda_t = \frac{S^0_t\,S^*_0}{S^*_t\,S^0_0}$ for $t \in [0,T]$. Note that $\Lambda_0=1$ and that when assuming that the putative
risk neutral measure $Q$ is an equivalent probability measure we get
 \begin{equation} 
\label{dt4'.b7a}
  E_t \left( \frac{\Lambda_T}{\Lambda_t}
 \,\frac{S^0_t}{S^0_T} \,H_T \right)  = E_t^Q \left( \frac{S^0_t}{S^0_T} \,H_T \right) ,
 \end{equation} 
where $\Lambda_t$ represents in a complete market the respective (Radon-Nikodym derivative)
density at time $t$ and $E^Q_t$ denotes conditional expectation under $Q$. We remark that $Q$ is an equivalent probability measure if and only if $\Lambda_t$ forms a true martingale, which means that the benchmarked savings account ${\hat{S}}^0_t$ forms a true martingale.

\begin{figure}
  
  \centering
    \includegraphics[width=\textwidth]{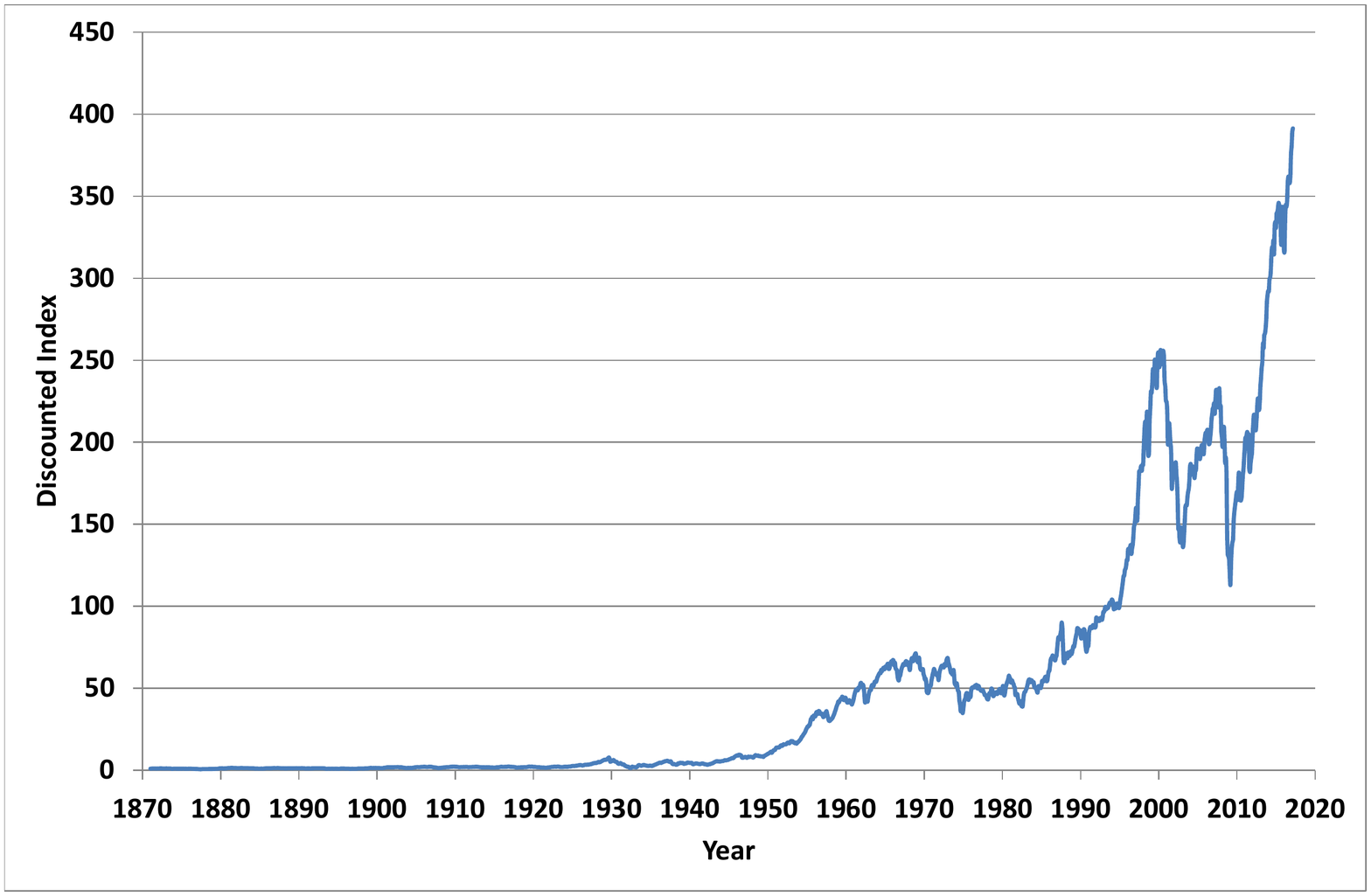}
	\caption{Discounted S$\&$P500 total return index.}    
	\label{fig:val41}
\end{figure}

\begin{figure}
  
  \centering
    \includegraphics[width=\textwidth]{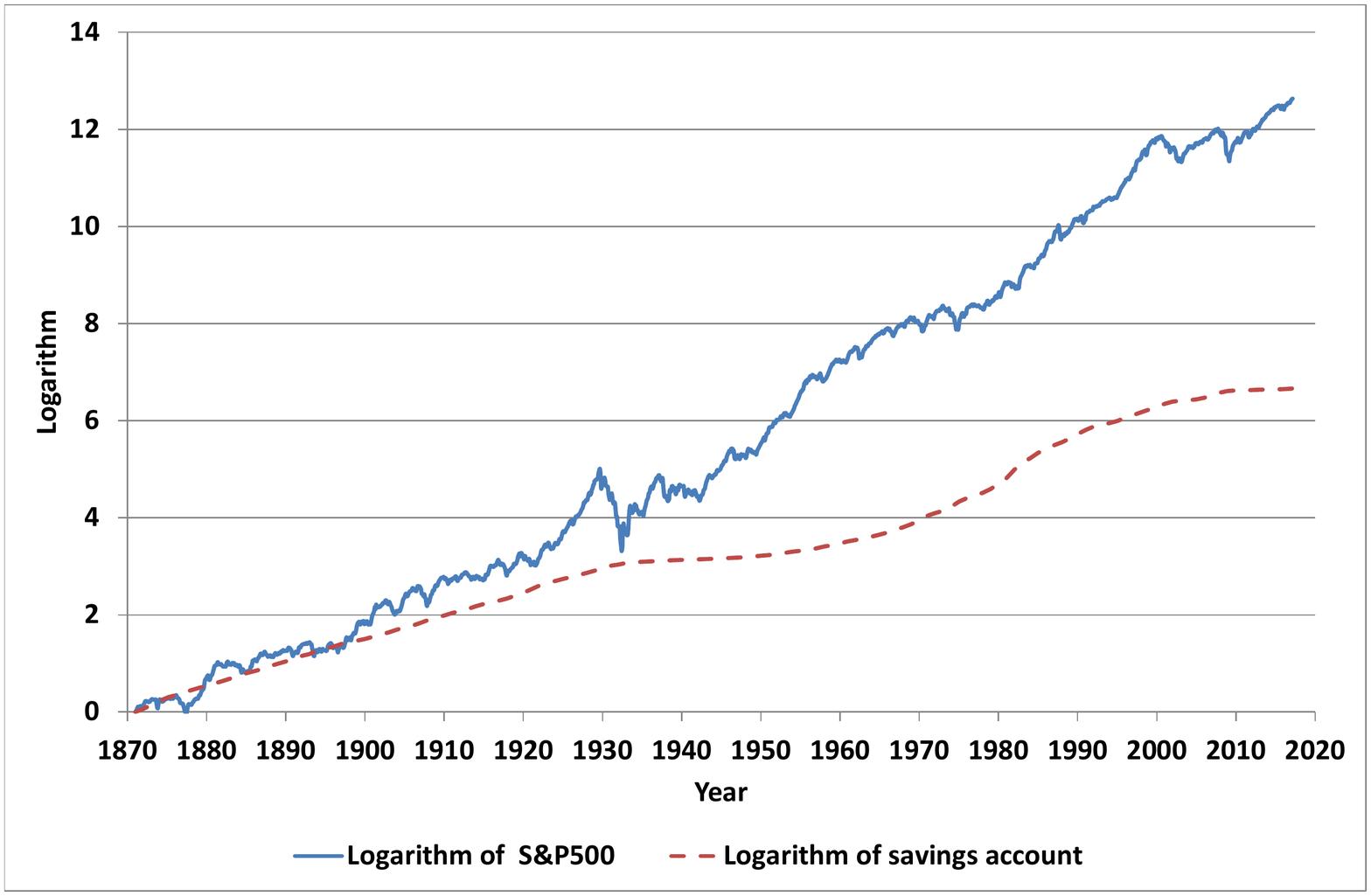}
	\caption{Logarithms of S$\&$P500 and savings account.}    
	\label{fig:val411}
\end{figure}

For illustration, let us interpret throughout this paper the S$\&$P500 total return
index as the benchmark and num\'eraire portfolio for the US equity
market. Its monthly observations in units of the US dollar savings
account are displayed in Figure~\ref{fig:val41} for the period from January
1871 until March 2017. The logarithms of the S$\&$P500 and the US dollar
saving account are exhibited in Figure~\ref{fig:val411}. One clearly
notes the higher long term growth rate of the S$\&$P500 when compared
with that of the savings account, a stylized empirical fact which is essential for the existence of the
stock market.

%

\begin{figure}
  
  \centering
    \includegraphics[width=\textwidth]{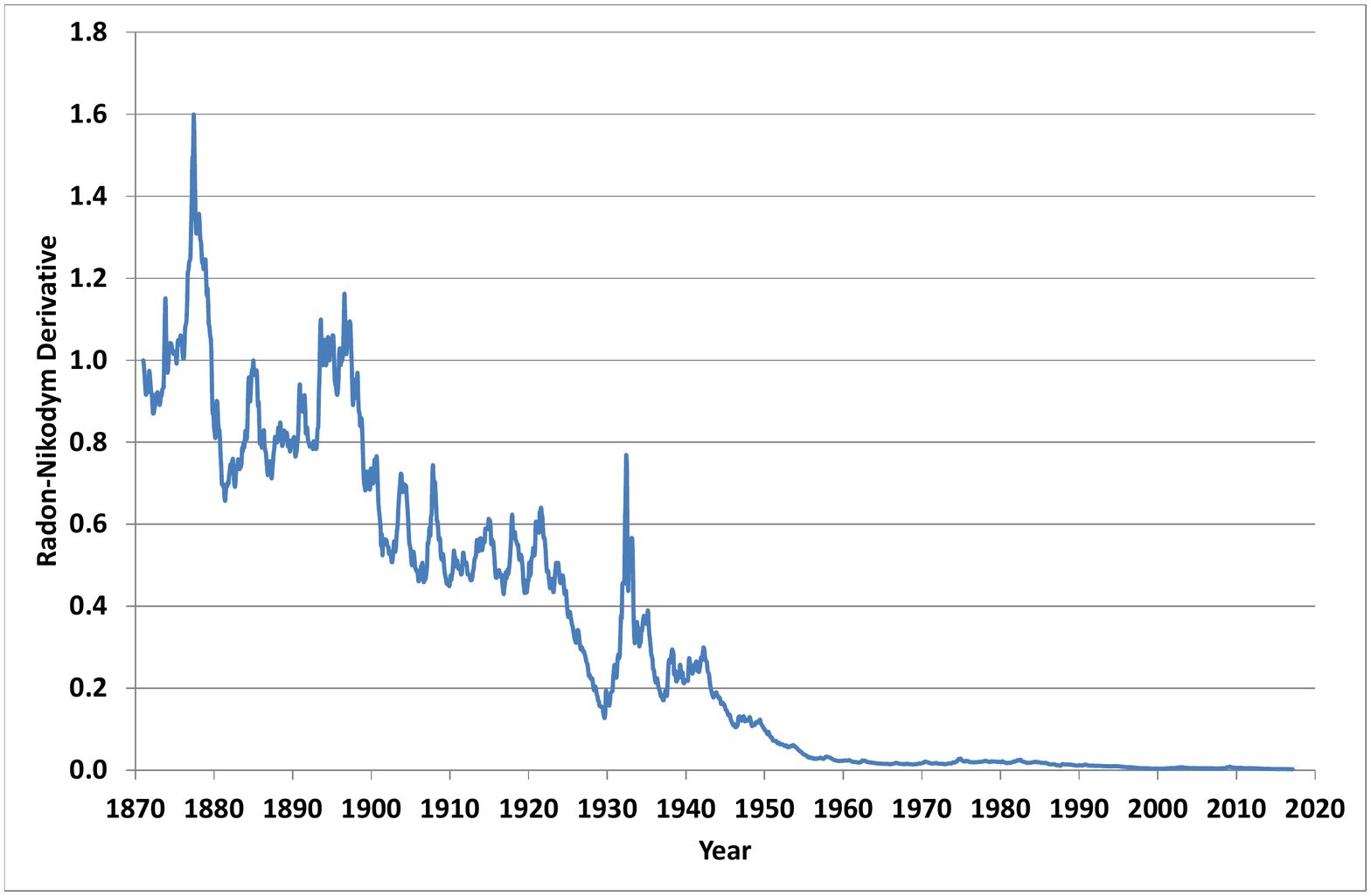}
	\caption{Radon-Nikodym derivative of the putative risk neutral measure in dependence on time $T$.}    
	\label{fig:val42}
\end{figure}
%
%
The normalized inverse of the discounted S$\&$P500
allows us to plot in Figure~\ref{fig:val42}
the resulting density process $\Lambda=\{\Lambda_t \, , \, t \in [0,T]\}$ of the putative risk neutral
measure $Q$ as it appears in \eqref{dt4'.b7}. Although one only has one sample path
to work with, it seems unlikely that the path displayed in
Figure~\ref{fig:val42} is the realization of a true martingale. Due to its
obvious systematic downward trend it seems
more likely to be the trajectory of a strict supermartingale.
In this case the density process would {\em not\/} describe a
probability measure and one could
expect substantial overpricing
to occur in risk neutral valuations of long term contracts when simply assuming that
the density process is a true martingale. Note that this martingale condition is the key
assumption of the theoretical foundation of classical risk neutral valuation, see
\cite{DelbaenSc98}. In current industry practice and most theoretical work this assumption is typically made. Instead of working on the filtered probability space $(\Omega , {\mathcal{F}}, {\underline{\mathcal{F}}}, P)$ one works on the filtered probability space
$(\Omega , {\mathcal{F}}, {\underline{\mathcal{F}}}, Q)$ assuming that there exists an equivalent risk neutral probability measure $Q$ without ensuring that this is indeed the case. In the case of a complete market we have seen that 
the benchmarked savings account has to be a martingale to ensure that risk neutral prices are
theoretically founded as intended. We observed in \eqref{dt4'.b7} and \eqref{dt4'.b7a}
that in this case the real world and the risk neutral valuation coincide. In the case when the benchmarked savings account is not a true martingale one can still perform formally risk neutral pricing. For a hedgeable nonnegative contingent claim one obtains then a self-financing hedge portfolio with values that represent the formally obtained risk neutral value. This portfolio when benchmarked is a supermartingale as a consequence of Assumption~\ref{ass:dt3.2b}.
Therefore, employing formally
obtained risk neutral prices by
the market
does not generate any
economically meaningful arbitrage in the sense as previously discussed.
However,  this may generate some classical form of arbitrage, as discussed in \cite{LoewensteinWi00} and \cite{Plat2002b}. Under the benchmark approach such classical forms of arbitrage are allowed to exist and can be systematically exploited, as we will demonstrate later on for the case of long dated bonds and similar contracts. The best performing portfolio is then still the benchmark or num\'eraire portfolio which remains finite at any finite time.

Finally, consider the valuation of nonhedgeable contingent claims.
Recall that the conditional expectation of a square integrable random variable can be interpreted as a
least squares projection; see
\cite{Shiryaev84}. Consequently, the real world valuation formula
\eqref{val3.3} provides with its conditional expectation, the
least squares projection of a given square integrable benchmarked payoff
into the set of possible current benchmarked values. It is well-known that in a
least squares projection the forecasting error has mean zero and
minimal variance, see \cite{Shiryaev84}. Therefore, the benchmarked
hedge error has mean zero and minimal variance. 
More predisely, as shown in \cite{DuPl16}, under benchmark risk minimization the Law
of the Minimal Price ensures 
through its real world valuation
that the value of the
contingent
claim is the minimal possible value and the benchmarked profit and loss has minimal fluctuations and is a local martingale orthogonal to all benchmarked traded wealth.

In an insurance company the benchmarked profits and losses
of diversified benchmarked contingent claims are pooled. If these benchmarked profits and losses are
generated by sufficiently independent sources of uncertainty,
then it follows intuitively via the Law of Large Numbers that the total benchmarked
profit and loss for an increasing number of benchmarked contingent claims is not only a local martingale starting at zero,
but also a process with an asymptotically vanishing quadratic variation or variance.
In this manner, insurance companies can
theoretically complete asymptotically the market by pooling
benchmarked profits and losses. This shows that real world valuation makes
perfect sense from the perspective of a financial
institution with a large pool of sufficiently different contingent claims.

\section{Valuation of Long Term Annuities}
 
  \label{sec:val8}
This section illustrates the real world valuation methodology
in the context of simple long term contracts, which we call here basic annuities.
More complicated annuities, life insurance products, pensions and also equity linked
long term contracts can be treated similarly.
All show, in general, a similar effect where real world prices become significantly lower
than prices formed under classical risk neutral valuation. The most important building blocks of annuities,
and also many other contracts, are zero coupon bonds. This section will, therefore, study first
in detail the valuation and hedging of zero coupon bonds. It will then
apply these findings to some basic annuity and compare its real world price with its
classical risk neutral price.

\subsection{Savings Bond} \label{subsec:val51}
To make the illustrations reasonably realistic, the following study
considers the US equity market as investment universe.
It uses the US one-year cash deposit rate as short rate when constructing
the savings account. The S$\&$P500 total return index is chosen as
proxy for the num\'eraire portfolio, the
benchmark. Monthly S$\&$P500 total return
data is sourced from Robert Shiller's website (http://www.econ.yale.edu/~shiller/data.htm) for
the period from January 1871 until March 2017. The savings account
discounted S$\&$P500 total return index has been already displayed in
Figure~\ref{fig:val41}.

For simplicity, and to make the core effect very clear, assume that the short rate is deterministic. By
making the short rate random one would complicate the
exposition, and would obtain very similar and even
slightly more pronounced differences between real world and risk neutral valuation, due to the effect of stochastic
interest rates on bond prices as a consequence of Jensen's
inequality. A similar comment applies to the choice of the
S$\&$P500 total return index as proxy for the num\'eraire portfolio or benchmark. Very
likely there exist
better proxies for the num\'eraire portfolio, see e.g. \cite{LePl06a} or \cite{PlRe17}.
As will become clear, their larger long term growth rates would make
the effects to be demonstrated even more pronounced.

\begin{figure}
  
  \centering
    \includegraphics[width=\textwidth]{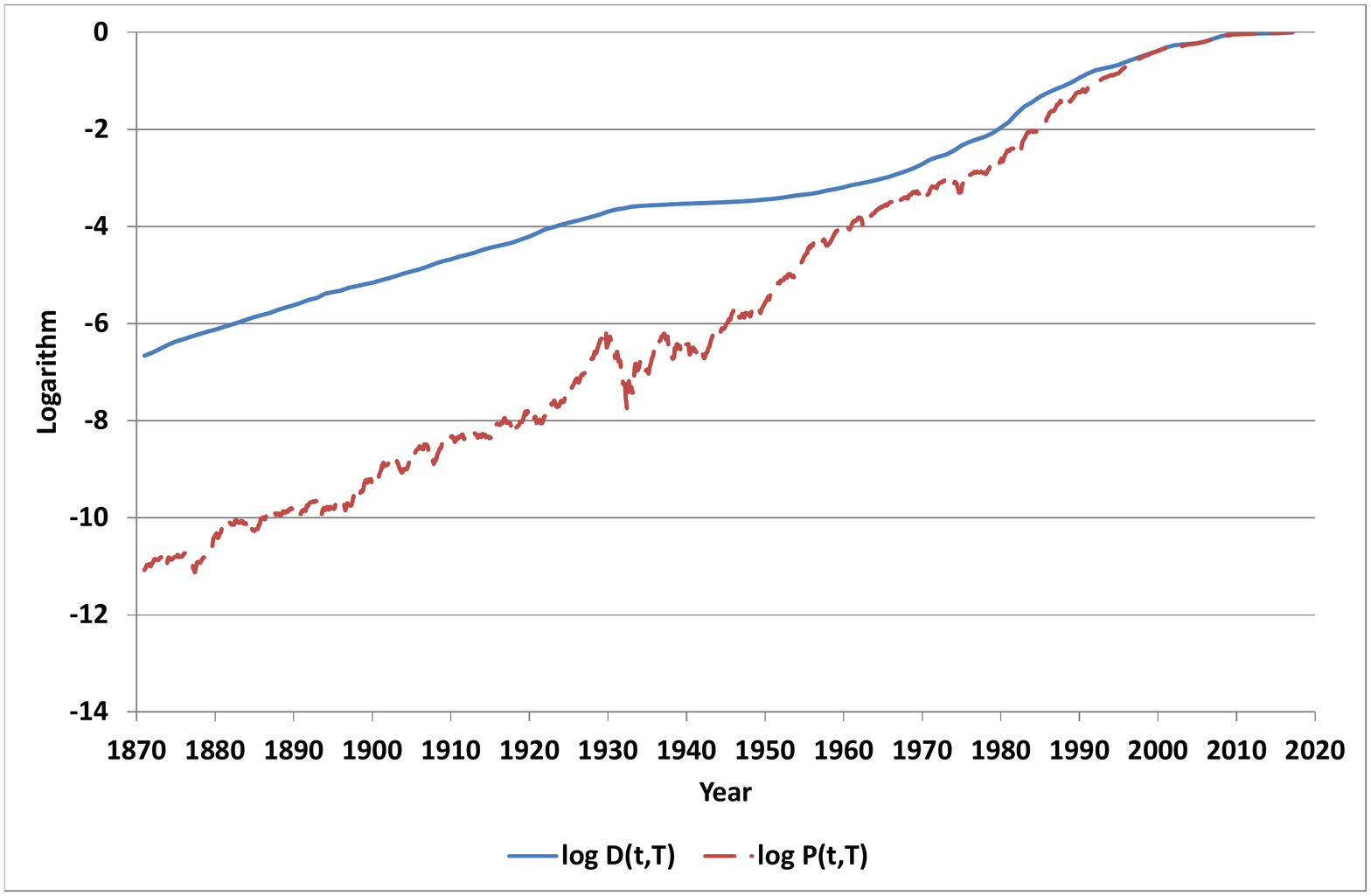}
	\caption{Logarithms of values of savings bond and fair zero coupon bond.}    
	\label{fig:val51}
\end{figure}

\begin{figure}
  
  \centering
    \includegraphics[width=\textwidth]{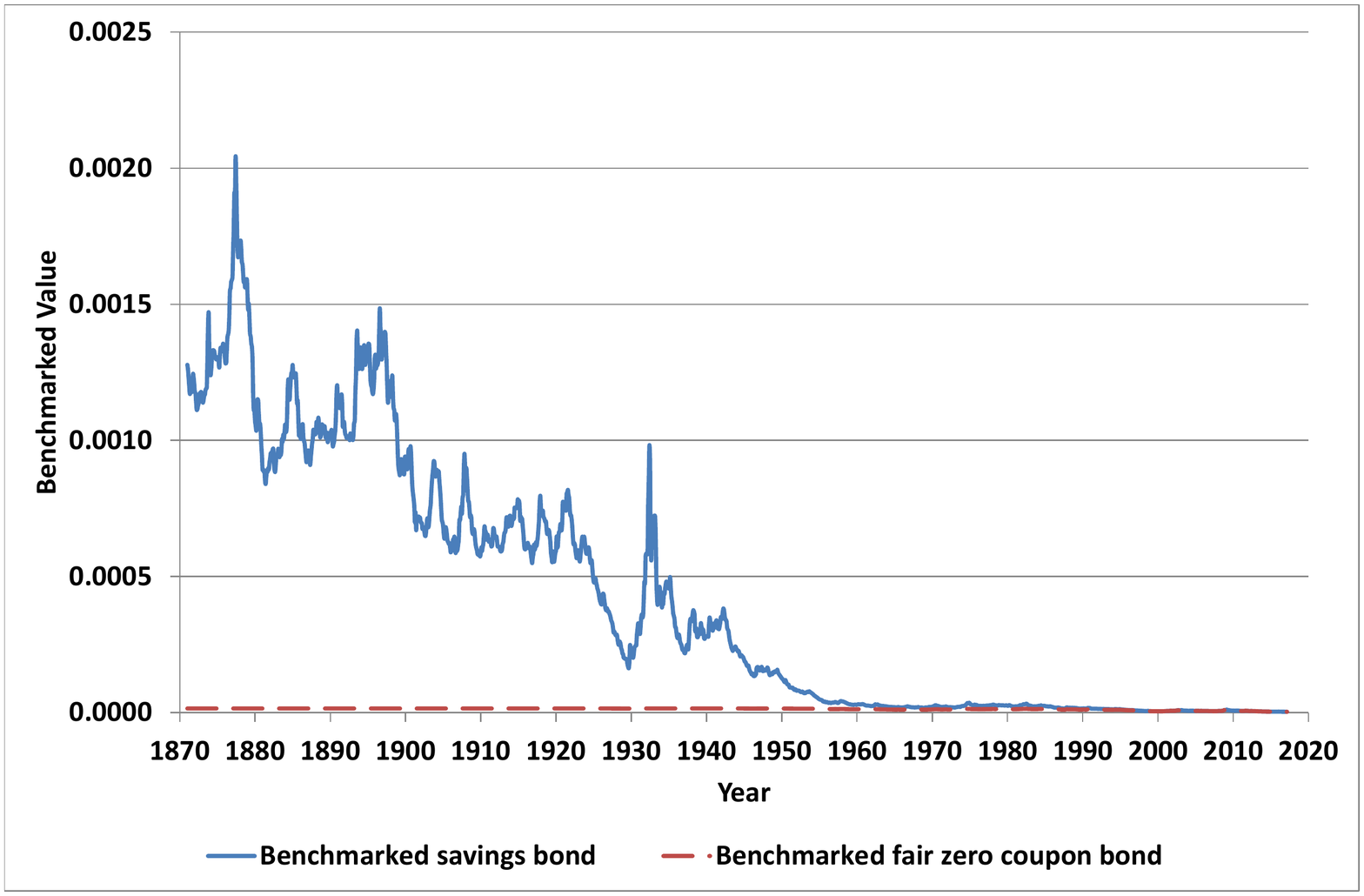}
	\caption{Values of benchmarked savings bond and benchmarked fair zero coupon bond.}    
	\label{fig:val52}
\end{figure}

The first aim of this section is to illustrate
the fact that under the benchmark approach
there may exist several self-financing portfolios that replicate
the payoff of one dollar of a zero coupon bond. Let
\begin{equation} 
 \label{subsec:val511}
 D(t,T)=\frac{S^0_t}{S^0_{T}}
 \end{equation} 
denote the price at time
$t \in [0,T]$, of the, so-called, {\em savings bond\/} with
maturity $T$, where $S^0_t$ denotes the value of the savings account at time $t$.
Consequently, the savings bond price is the price of a zero coupon
bond under risk neutral valuation and other classical valuation approaches.
The upper graph in Figure~\ref{fig:val51} exhibits the logarithm of the saving bond
price, with maturity in March 2017 valued at the time shown on the x-axis. The benchmarked value
of this savings bond,
which equals its value denominated in units of the S$\&$P500, is displayed
as the upper graph in Figure~\ref{fig:val52}.

%
As pointed out previously, an equivalent risk neutral probability measure is
unlikely to exist for any realistic complete market model.
Financial planning recommends
to invest at young age in the equity market and to shift wealth into fixed income securities closer
to retirement. This type of strategy is widely acknowledged to be more efficient than investing all wealth
in a savings bond, which represents the classical risk neutral strategy for obtaining bond payoffs at maturity. 
Based on the absence of an equivalent risk neutral probability measure, this paper provides a theoretical reasoning for such a long term investment strategy.
Moreover, it
will quantify rigorously such a strategy under the assumption of a stylized model for the benchmark dynamics.

\subsection{Fair Zero Coupon Bond}

Under real world valuation, 
the time $t$ value of the {\em fair zero coupon bond price}
with maturity date $T$ is denoted by
 \begin{equation} 
 \label{lm2.8}
 P(t,T) = S^*_t\,E_t\left( \frac{1}{S^*_T}\right) ,
 \end{equation} 
and results from the real world valuation formula \eqref{val3.3}, see
also \eqref{dt4'.63b}. It provides the minimal possible price for a
self-financing portfolio that replicates $\$1$ at maturity $T$. Note that
under real world valuation the fair zero coupon bond becomes an
index derivative, with the index as benchmark. The underlying assets involved, are the benchmark (here
the S$\&$P500 total return index) and the savings account of the domestic currency, here the US dollar. Both securities
will appear in the corresponding hedge portfolio, which shall replicate at maturity the payoff
of one dollar.

To calculate the price of a fair zero coupon bond, one has to
compute the real world conditional expectation in \eqref{lm2.8}. For this
calculation one needs to employ a model for the real world
distribution of the random variable $(S^*_T)^{-1}$.
To be realistic and different to the formally obtained risk neutral valuation, such a model must reflect the fact that
the benchmarked savings account should be a strict supermartingale. 
Any model that models the benchmarked savings account as a strict supermartingale,
will value the above
benchmarked fair zero coupon bond 
less expensively than the
corresponding benchmarked savings bond.
Figure~\ref{fig:val52} displays, additionally to the benchmarked savings bond, the value of a benchmarked fair zero coupon bond,
which will be derived below, under a respective model. One notes the significantly lower initial value of the benchmarked fair
zero coupon bond.
Also visually, its benchmarked value seems to appear as the best forecast of its future benchmarked values.
In Figure~\ref{fig:val51} the logarithm of the fair zero coupon bond is shown together with the logarithm of the
savings bond value. One notes that the fair zero coupon bond appears to follow essentially the benchmark for many years.
The strategy that delivers this hedging porfolio will be discussed in detail below.

We interpret the value of the savings bond
as the one obtained by formal risk neutral valuation.
As shown in relation \eqref{dt4'.b7a}, this value is greater than or equal to that of the fair
zero coupon bond. For readers who want to have some economic
explanation for the observed value difference one could argue that the
savings bond gives the holder the right
to liquidate the contract at any time without costs. On the other hand, a fair zero
coupon bond is akin to a term deposit without the right to access
the assets before maturity. One could say that the savings bond
carries a ``liquidity premium" on top of
the value for the fair zero coupon bond.
Under the classical no-arbitrage paradigm, with its Law of One Price (see e.g. \cite{Tayl02}), there is only
one and the same price process possible for both instruments, which is that of the savings bond.
The benchmark approach opens with its real world valuation concept the possibility to model
costs for early liquidation of financial instruments.
The fair zero coupon bond is the least liquid instrument that delivers the bond payoff
and, therefore, the least expensive zero
coupon bond. The savings bond is more liquid and, therefore, more expensive.

\subsection{Fair Zero Coupon Bond for the Minimal Market Model}

The benchmarked fair zero coupon value at time $t\in [0,T]$ is the best forecast of its
benchmarked payoff ${\hat{S}}^0_T = (S^*_T)^{-1}$. It provides the minimal self-financing portfolio value process that hedges
this benchmarked contingent claim. To facilitate a tractable
evaluation of a fair zero coupon bond, one has to employ a
continuous time model for the benchmarked savings
account that represents a strict supermartingale.
The inverse of the benchmarked savings account is the
discounted num\'eraire portfolio ${\bar{S}}^*_t=\frac{S^*_t}{S^0_t} = ({\hat{S}}^0_t)^{-1}$. In the illustrative example we present, it is
the discounted S$\&$P500, which, as discounted num\'eraire portfolio,
satisfies in a continuous market model the stochastic differential
equation (SDE)
 \begin{equation} 
 \label{lm2.8'}
 d\, {\bar{S}}^*_t =\alpha_t\,dt + \sqrt{{\bar{S}}^*_t\,\alpha_t}
 \,dW_t,
 \end{equation} 
for $t\ge 0$ with ${\bar{S}}^*_0 >0$, see \cite{PlHe2010} Formula (13.1.6). Here
$W=\{W_t, t\ge 0\}$ is a Wiener process, and $\alpha=\{\alpha_t,
t\ge 0\}$ is a strictly positive process, which models the trend
$\alpha_t={\bar{S}}^*_t \theta^2_t$ of ${\bar{S}}^*_t$, with $\theta_t$
denoting the market price of risk. For constant market price of risk one would obtain the Black-Scholes model, which has been the standard market model. In the long term it yields benchmarked savings accounts that are martingales and is, therefore, not suitable for our study. 
Since in \eqref{lm2.8'} $\alpha_t$ can be a rather general stochastic process,
the parametrization of the SDE \eqref{lm2.8'}
 does so far not constitute a model.

The trend or drift in the SDE \eqref{lm2.8'} can be interpreted economically as a measure for the discounted average value of
wealth generated per unit of time by the underlying economy.
To construct in a first approximation a respective model, we assume now that
the drift of the discounted S$\&$P500 total return index grows exponentially
with a net growth rate $\eta > 0$.
At time $t$ the drift of the discounted S$\&$P500 total return
index ${\bar{S}}^*_t$ is then modeled by the exponential function
\begin{equation} 
\alpha_t= {\alpha}\,\exp\{\eta\,t\}.
\end{equation} 
This yields the stylized version of the
{\em minimal market model\/} (MMM),
see \cite{Platen01a,Plat2002b}, which emerges from \eqref{lm2.8'} and is the `workhorse' of the benchmark approach.
We know explicitly the transition density of the resulting time-transformed
squared Bessel process of dimension four,
${\bar{S}}^*$, see \cite{RevuzYo99}. 

The transition density function of the discounted num\'eraire portfolio ${\bar{S}}^*$ equals
\begin{equation}
\label{Eqn:transitiondensityMMMdiscountedGOP}
p_{{\bar{S}}^*}(t, x_t , {\bar{T}} , x_{\bar{T}} ) = \frac{1}{2(\varphi_{\bar{T}} - \varphi_t )} \sqrt{\frac{x_{\bar{T}}}{x_t}} \exp \left( -\frac{x_t + x_{\bar{T}}}{2(\varphi_{\bar{T}} - \varphi_t )} \right) I_1 \left( \frac{\sqrt{x_t x_{\bar{T}}}}{\varphi_{\bar{T}} - \varphi_t} \right) ,
\end{equation}
where $\varphi_t = \frac{1}{4\eta }\alpha (\exp(\eta t)-1)$ is also the quadratic variation of $ \sqrt{{\bar{S}}^*} $.
The corresponding distribution function is that of a non-central chi-squared random variable with four degrees of freedom and non-centrality parameter $x_t/(\varphi_T - \varphi_t)$.
Using the above transition density function we apply standard maximum likelihood estimation to monthly data for the discounted S$\&$P500 total
return index over the period from January 1871 to January 1932, giving the following estimates of the parameters $\alpha$ and $\eta$,
\begin{align}
\label{Eqn:ParaEstMMM}
\alpha &= \MMMalphabar\, (\MMMalphabarSE ), \\ \notag
\eta &= \MMMeta \, (\MMMetaSE) ,\end{align}
where the standard errors are shown in brackets. In Appendix~\ref{App:A} we explain the estimation used in this paper.

These estimates for the net growth rate $\eta$ are consistent
with estimates from various other sources in the literature, where the net growth rate of
the US equity market during the last
century has been estimated at about $5\%$, see for instance
\cite{DimsonMaSt02}. 

Under the stylized MMM the explicitly known transition density of the discounted
num\'eraire portfolio ${\bar{S}}^*_t$ yields
for the fair zero coupon bond price by \eqref{lm2.8} the explicit
formula
 \begin{equation} \label{dt5'.4}
 P(t,T) =
 D(t,T)\left( 1-\exp\left\{-\frac{2\,\eta\,{\bar{S}}^*_t}
 {\alpha\,(\exp\{\eta\,T\} - \exp\{\eta\,t\})} \right\}\right) 
 \end{equation} 
for $t \in [0,T)$, which has been first pointed out in 
\cite{PlHe2010} Section 13.3.
Figure~\ref{fig:val52} displays with the lower graph the trajectory
of the benchmarked fair zero coupon bond price with maturity $T$
in March 2017. By \eqref{dt5'.4} the price of the benchmarked fair
zero coupon bond remains always below that of the benchmarked
savings bond, where the latter we interpret as the formally taken risk neutral
zero coupon bond price. The fair zero coupon
bond value provides the minimal portfolio process for hedging
the given payoff under the assumed MMM. Other benchmarked replicating portfolios
need to form strict supermartingales and, therefore, yield higher price processes. One such example is
given by the benchmarked savings bond, which pays one dollar at maturity. Recall that the benchmarked
fair zero coupon bond is a martingale. It is minimal
among the supermartingales that represent benchmarked replicating self-financing
portfolios and pay one dollar at maturity.

\begin{figure}
  
  \centering
    \includegraphics[width=\textwidth]{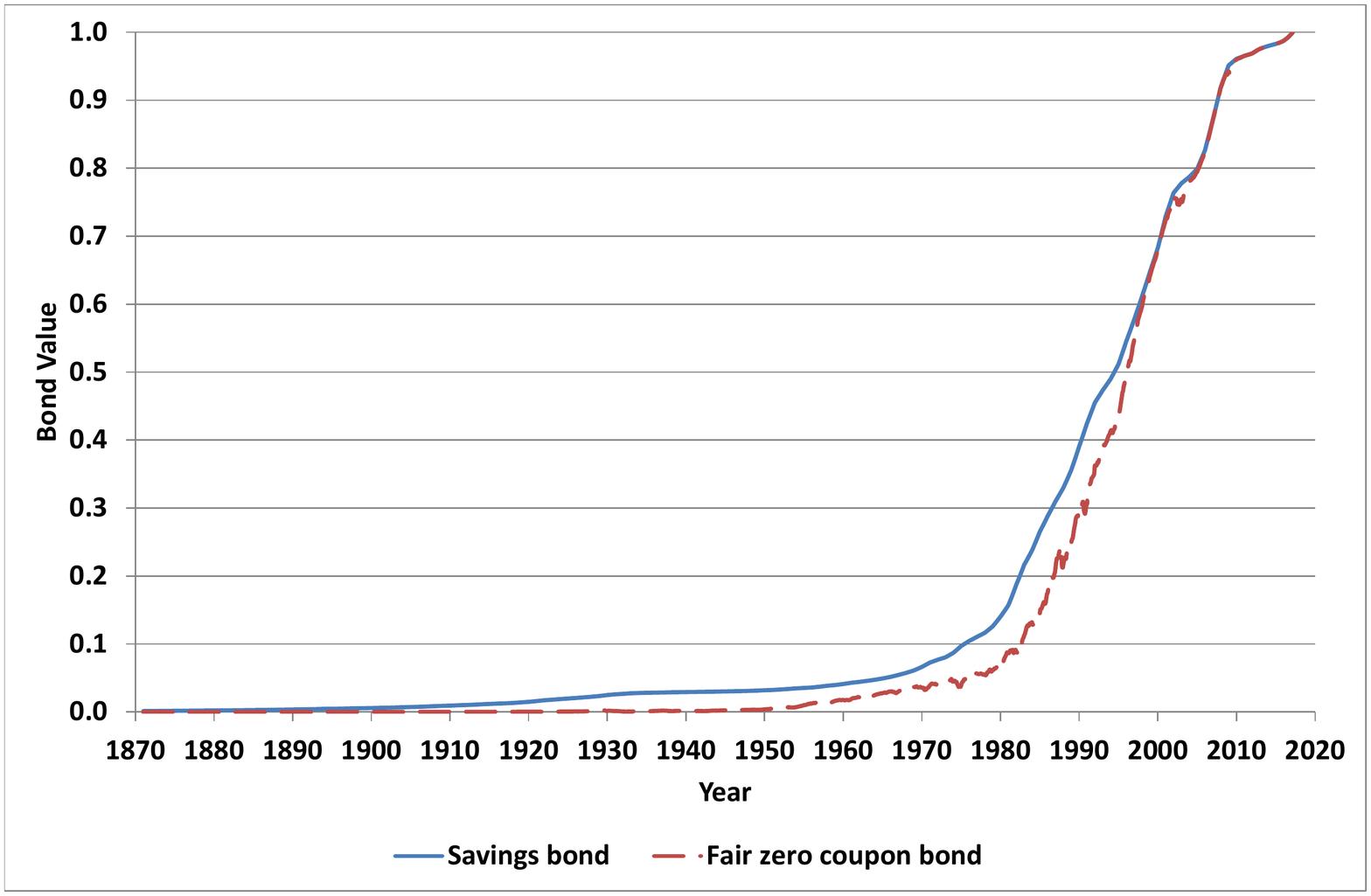}
	\caption{Values of savings bond and fair zero coupon bond.}    
	\label{fig:val53}
\end{figure}

Figure~\ref{fig:val53} exhibits with its upper graph the trajectory
of the savings bond and with its lower graph that of the
fair zero coupon bond in US dollar denomination. Closer to
maturity the fair zero coupon bond price merges with the
savings bond price. Both self-financing portfolios
replicate the payoff at maturity. 
Most important is the observation that they start with
significantly different initial prices.
The fair zero coupon bond exploits the presence of the strict supermartingale property of the benchmarked savings account,
whereas the savings bond ignores it totally.
Two
self-financing replicating portfolios are displayed in Figure~\ref{fig:val53}.
Such a situation, where two self-financing portfolios replicate the same contingent claim, is
impossible to model under the classical no-arbitrage
paradigm. However, under the benchmark approach this is a natural situation.

In the above example for the parameters estimated during the period from January 1871 to January 1932, the savings bond with maturity in March 2017 has
in January 1932 a price of $D(0,T) \approx \$0.026596$. The fair
zero coupon bond is far less expensive and priced at only $P(0,T)
\approx \$0.000709$. The fair zero coupon bond with term to
maturity of more than 80 years costs here less than $3\%$ of the savings
bond. This reveals a substantial premium
in the value of the savings bond.

We repeated with the estimated parameters the study for all possible zero coupon bonds that cover a period of 10, 15, 20 and 25 years from initiation until maturity that fall into the period starting in January 1932 and ending in March 2017.
Table~\ref{Tab:1} displays the average difference in US dollars between the risk neutral and the fair bond. One clearly notes that for a 25-year bond one saves about 27\% of the risk neutral price, which is a typical time to maturity for many pension products.

\begin{table}
\begin{tabular}{cccc}
Years to Maturity & Mean of D(t,T) & Mean of P(t,T) & Mean of \{D(t,T)-P(t,T)\}\\
\hline
10 & 0.6534 & 0.6468 & 0.0066\\
15 & 0.5219 & 0.4931 & 0.0287\\
20 & 0.4084 & 0.3479 & 0.0604\\
25 & 0.3130 & 0.2294 & 0.0836\\
\end{tabular}
\caption{Mean values of zero coupon bonds of prescribed terms to maturity whose start and end dates lie between January 1932 and March 2017.}
\label{Tab:1}
\end{table}

By demonstrating explicitly the valuation methodology we suggest a realistic way of changing from the classical risk neutral production strategy to the less expensive benchmark production strategy.

\subsection{Hedging of a Fair Zero Coupon Bond}

The benefits of the proposed benchmark production methodology can only be harvested if the respective hedging strategy would allow to replicate the hedge portfolio values as theoretically predicted.

The hedging strategy by
which this is theoretically achieved follows under the MMM from the explicit fair zero coupon bond pricing formula \eqref{lm2.8'}.
At the time $t \in [0,T)$
the corresponding theoretical number of units of the S$\&$P500 to
be held in the hedge portfolio follows, similar to the well-known Black-Scholes delta hedge ratio, 
as a partial derivative with respect to the underlying, 
and is given by the formula
 \begin{eqnarray}
 \label{dt5'.5}
 \delta^*_t &=& \frac{\partial {\bar{P}}(t,T)}{\partial
 {\bar{S}}^*_t} \nonumber \\
 &=& D(0,T)\, \exp\left\{\frac{- 2\,\eta\,{\bar{S}}^*_t}
 {\alpha\,(\exp\{\eta\,T\} - \exp\{\eta\,t\})} \right\} \,
 \frac{2\,\eta}
 {\alpha\,(\exp\{\eta\,T\} - \exp\{\eta\,t\})}.
 \end{eqnarray}
Here $D(0,T)$ is the respective value of the formally obtained risk neutral bond, the savings bond.

\begin{figure}
  
  \centering
    \includegraphics[width=\textwidth]{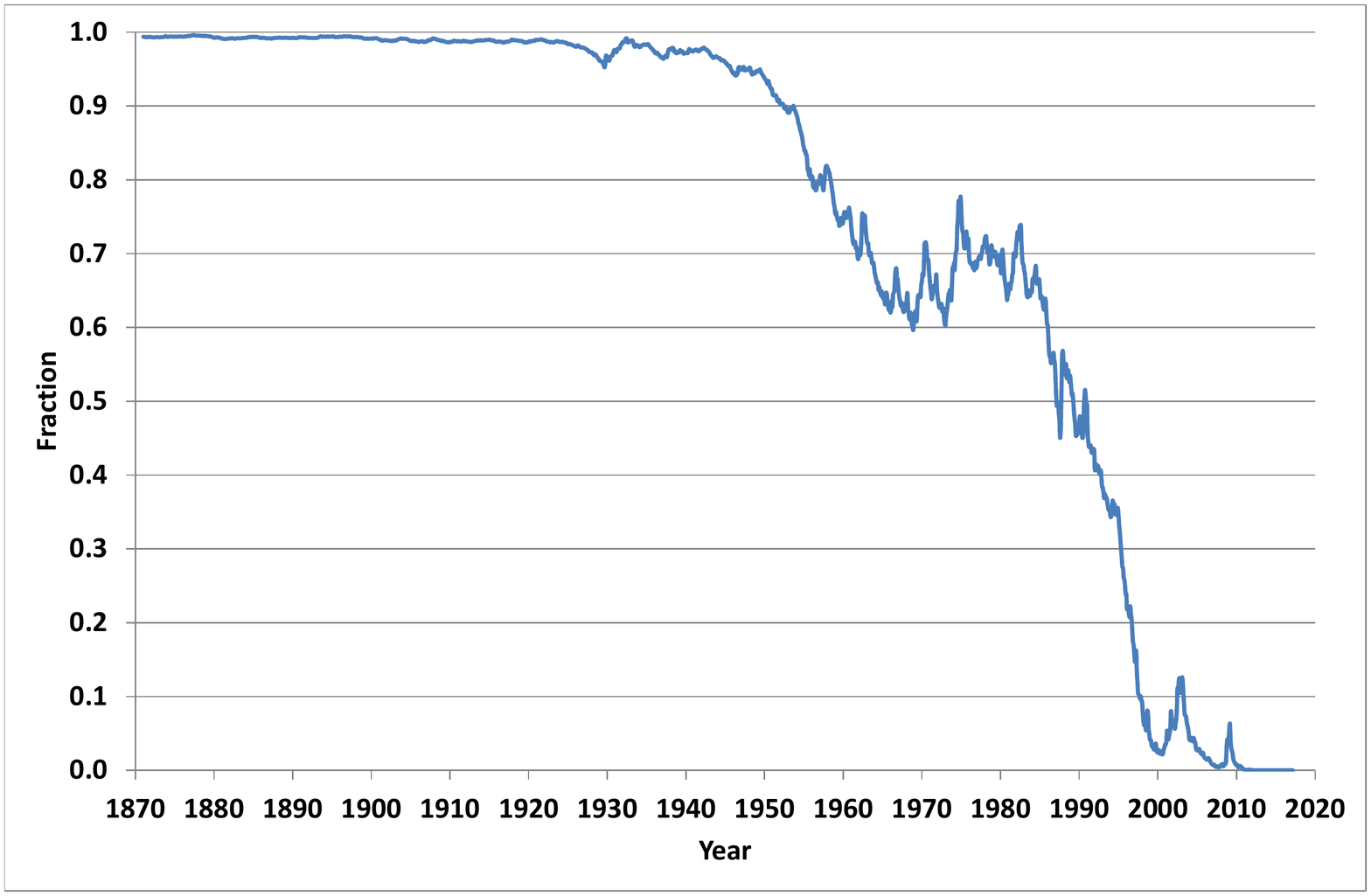}
	\caption{Fraction of wealth invested in the S$\&$P500.}    
	\label{fig:val54}
\end{figure}

The resulting fraction of wealth to be held
at time $t$ in the S$\&$P500, as it evolves for the given example of zero coupon bond valuation from January 1932 to maturity, shown in Figure~\ref{fig:val54}. The remaining wealth is always invested in the
savings account.

\begin{figure}
  
  \centering
    \includegraphics[width=\textwidth]{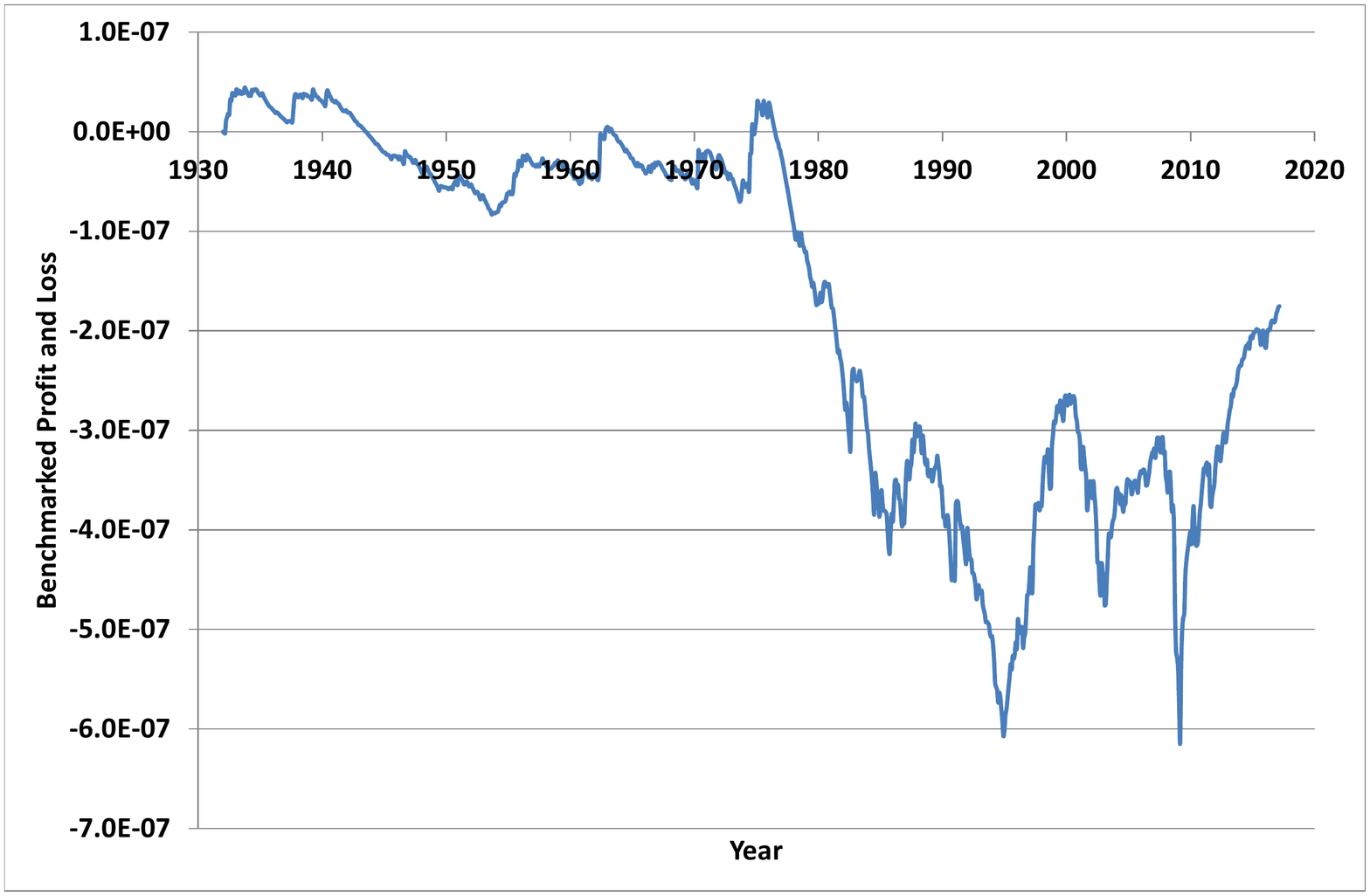}
	\caption{Benchmarked profit and loss.}    
	\label{fig:val55}
\end{figure}

\begin{figure}
  
  \centering
    \includegraphics[width=\textwidth]{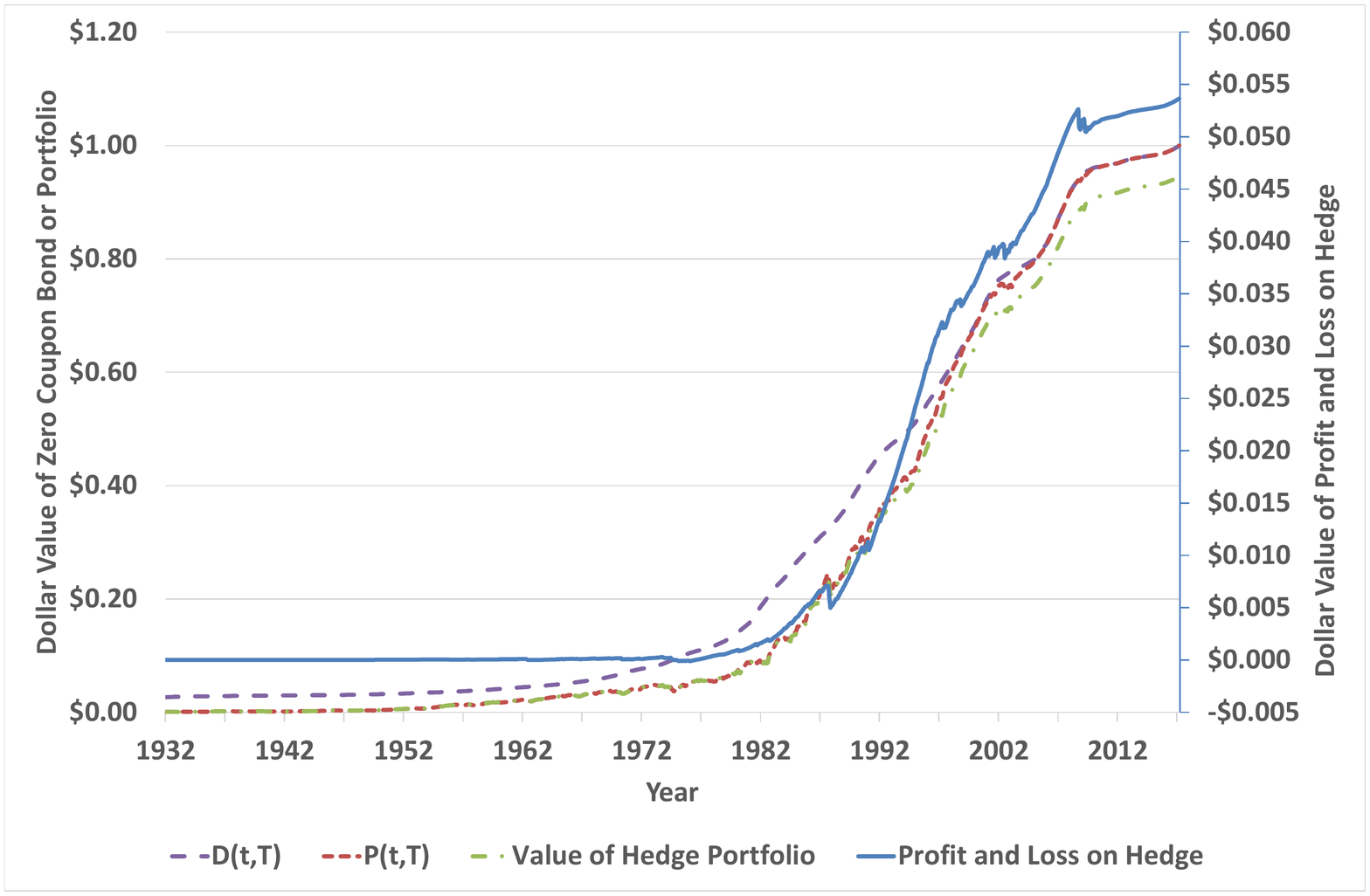}
	\caption{Comparison of values of zero coupon bonds with hedge portfolio, also showing the profit and loss.}    
	\label{fig:val55a}
\end{figure}

To demonstrate how realistic the hedge of the fair zero coupon
bond payoff is for the given delta under the stylized MMM under monthly reallocation,
a self-financing hedge portfolio is formed. The delta hedge is performed similarly to the well known
one for options under the Black-Scholes model. The
self-financing hedge portfolio starts in January 1932, which ensures that the hedge simulations employing the fitted parameters in \eqref{Eqn:ParaEstMMM} are out-of-sample. Each month
the fraction invested in the S$\&$P500 is adjusted in a
self-financing manner according to the above prescription. The
resulting benchmarked profit and loss for this
delta hedge turns out to be very small and is visualized in
Figure~\ref{fig:val55}. The maximum absolute benchmarked profit and loss
amounts only to about 0.00000061. This benchmarked profit and loss is so small that the
resulting hedge portfolio, when plotted additionally in
Figure~\ref{fig:val53}, would be visually indistinguishable from the path
of the already displayed fair 
zero coupon bond price process. Dollar values of the self-financing hedge portfolio, the fair zero coupon bond and the savings bond are shown in Figure~\ref{fig:val55a} where it is evident that the self-financing portfolio replicates 95\% of the face value of the bond but employs less than 3\% of the initial capital.
One may argue that this represents only one hedge simulation. Therefore, we employed the same 10, 15, 20 and 25 year fair bonds that fit from initiation until maturity into the period from January 1932 until March 2017 and performed analogous hedge simulations.
In Table~\ref{Tab:2} we report in 
US dollars
the average profits and losses with respective standard deviations, where for a 25-year ZCB the hedge losses at maturity average 2.26\% of the face value while the initial hedge portfolio is 27\% less expensive than the savings bond.

\begin{table}
\centering
\begin{tabular}{ccc}
Years to Maturity & Mean P\&L & Std. Dev. P\&L\\
\hline
10 & -0.0004 & 0.0078\\
15 & -0.0057 & 0.0108\\
20 & -0.0144 & 0.0138\\
25 & -0.0226 & 0.0168\\
\end{tabular}
\caption{Means and standard deviations of profits and losses on hedge at maturities of zero coupon bonds having various terms to maturity. (Negative P\&Ls indicate losses).}
\label{Tab:2}
\end{table}

The above example and hedge simulations demonstrate the principle that a hedge portfolio can generate a fair zero
coupon bond value process by investing long only in a dynamic hedge in the S$\&$P500
and the savings account. The resulting hedge portfolio is for long term maturities significantly
less expensive than the corresponding saving bond.
Moreover, as can be seen in Figure~\ref{fig:val51}, the hedge portfolio can initially significantly
fluctuate, as it did around 1930 during the Great Depression. However, close to maturity,
the hedge portfolio cannot be significantly affected by any equity market meltdown, as was the case in our illustrative example during the 2007-2008 financial crisis. 

As we will describe later on, in a similar manner as above described, one can value and produce less expensively other long term contingent claims including
equity linked payoffs that are typical for variable annuities or life insurance and pension payoffs.

In summary, one can say that by shifting the valuation paradigm from classical risk
neutral to real world valuation, we suggest to replicate more cost
efficiently long term payoffs. In particular, it follows from \eqref{dt4'.b7} that
one can expect significant savings to arise for payoffs that
do not vanish when the benchmark approaches zero.
This production methodology applies to a range of payoffs that are
embedded in various insurance and pension contracts, where we aim below to provide some indications.

\subsection{Long Term Mortality and Cash-Linked Annuities}
Consider now a stylized example that aims to illustrate valuation and hedging under the benchmark approach
in the context of basic annuities. Consider annuities sold to $K$ policy holders that pay an indexed number of units of the savings account per year at the beginning of each year where they provide a payoff. Here we assume that the savings account and total return index account commence with \$1 at the date $t_0$ of purchasing the annuity and the indexed number of units at time $T$ is prescribed as
\begin{equation}
MI_T = \frac{1}{1+\sum_{k=1}^K 1_{\tau^{(k)}>T }},
\end{equation}
where $\tau^{(k)}$ is a random variable equal to the time at which the $k$-th policy holder dies, for $k=1,2,\ldots , K$.
Therefore, the payoff at time $T$ in respect of the $k$-th policy holder is
\begin{equation}
MI_T\times \frac{S^0_T}{S^0_{t_0}} \times 1_{\tau^{(k)}>T }.
\end{equation}
Additionally, we assume that there is an asset management fee payable as an annuity whose payoff at time $T$ equals $MI_T\times \frac{S^0_T}{S^0_{t_0}}$.
This type of payoff is likely to account well in the long run for the effect of inflation because the average US interest rate was during the last century on average about 1\% above the US average inflation rate, see \cite{DimsonMaSt02}. Since in our evaluation the interest rate will not play any role, we can now assume that the interest rate is stochastic. Also, since the portfolio of annuities has at time $T$ the aggregate payoff 
\begin{equation}
MI_T\times \frac{S^0_T}{S^0_{t_0}} + \sum_{k=1}^K \bigg\{ MI_T\times \frac{S^0_T}{S^0_{t_0}} \times 1_{\tau^{(k)}>T } \bigg\} = \frac{S^0_T}{S^0_{t_0}} ,
\end{equation}
the mortality rate plays no role and we can assume that the mortality rate is stochastic.

To use the available historical data efficiently, let us place our discussion in the past and consider a person who reached the age of 25 in January 1932. The subset of the monthly S$\&$P500 time series used for fitting the parameters in \eqref{Eqn:ParaEstMMM} starts at January 1871 and ends at this date. The person may have considered purchasing some annuity which pays, from the age of 65 to the beginning of the age of 110, at the beginning of the year $T$
an indexed number of units $MI_T$ of the savings account. In the case when the person may have passed away before reaching the age of 110 in 2017 the payments that would have otherwise been made revert to the asset pool backing the annuity portfolio. 
Since in this setting there is no mortality risk or interest rate risk involved in the given payoff stream, classical risk neutral pricing would value this annuity portfolio as
\begin{equation}
V^{\textrm{RN}}_t = \sum_{T\in G} S^0_t E_t \left( \frac{S^0_T/S^0_{t_0}}{S^0_T} \right) = 45\times \frac{S^0_t}{S^0_{t_0}} 
\end{equation}
for the set of payment dates $G = \{\textrm{Jan }1972, \textrm{Jan }1973,\dots, \textrm{Jan }2016\}$.
Thus for any date $t$ during the period from January 1932 until December 1971 classical risk neutral pricing would always value this annuity as being equal to 45 units of the savings account. This is exactly the number of units of the savings account that have to be paid out by the annuity over the 45 years from 1972 until 2017. This means, the annuity has the discounted risk neutral value
\begin{equation} 
 \label{vrn}
 \bar{V}^{\textrm{RN}}_t=\frac{V^{\textrm{RN}}_t}{S^0_t}=45
\end{equation} 
for all $t \in \{\textrm{Jan }1932, \textrm{Feb }1932,\dots, \textrm{Dec }1971\}$.

\begin{figure}
  
  \centering
    \includegraphics[width=\textwidth]{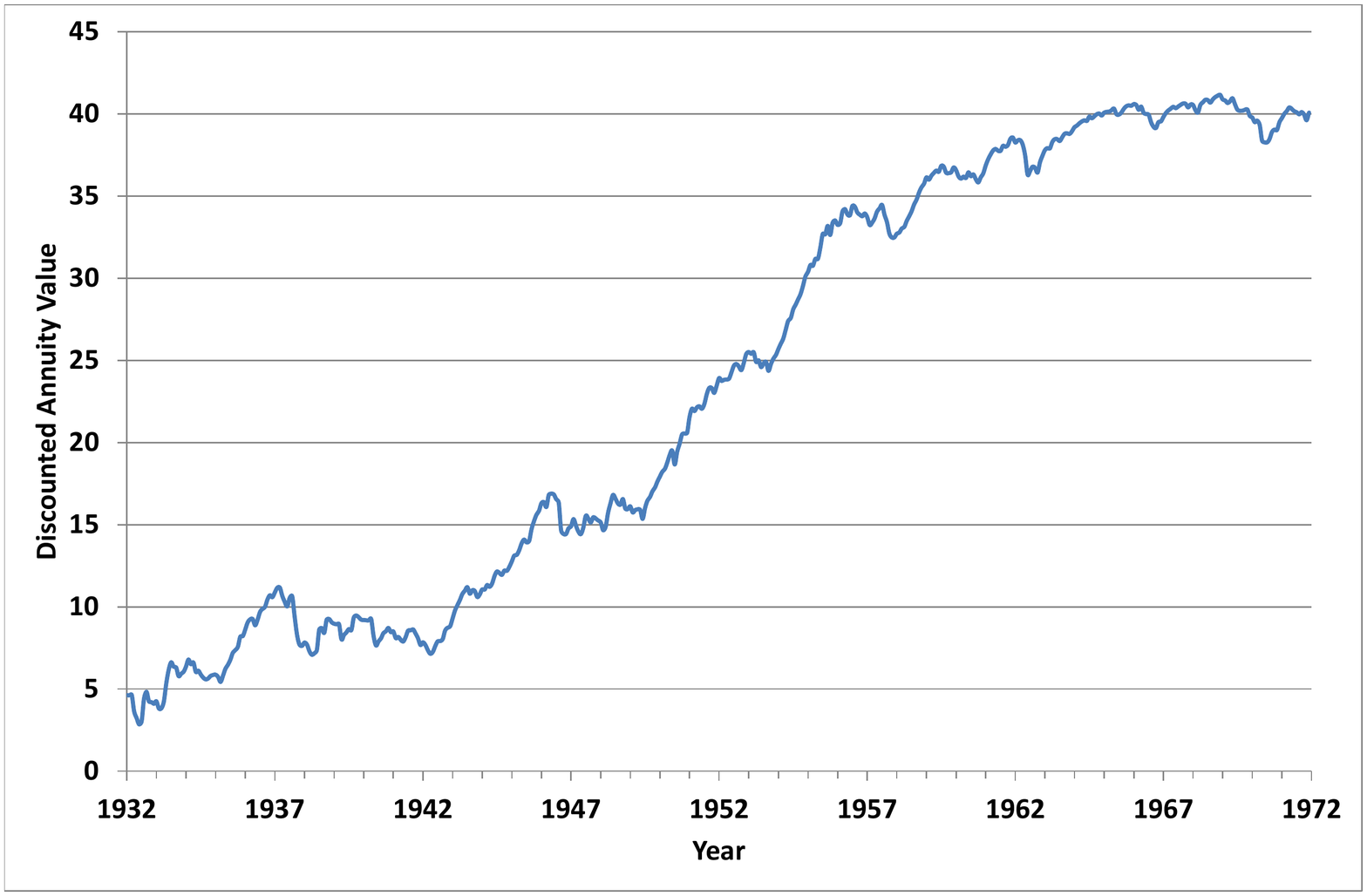}
	\caption{Savings account discounted price of the annuity during the period from January 1932 until December 1971 which provides annual payments from January 1972 until January 2016.}    
	\label{fig:val56}
\end{figure}

As shown in the previous section, when valuing such an annuity under the benchmark approach it will be less expensive than suggested by the above classical risk neutral price. This example aims to illustrate that significant amounts can be saved. The real world valuation formula, given in \eqref{val3.3}, captures at time $t$ the fair value of one unit of the savings account at time $T$, see \eqref{subsec:val511} and \eqref{lm2.8'}, via the expression
\begin{equation} 
 \label{val532}
S^*_t E_t\Big(\frac{S^0_T}{S^*_T}\Big)=
S^0_t E_t\Big(\frac{{\bar S}^*_t}{{\bar S}^*_T}\Big) = S^0_t\left( 1-\exp\left\{-\frac{2\,\eta\,{\bar{S}}^*_t}
 {\alpha\,(\exp\{\eta\,T\} - \exp\{\eta\,t\})} \right\}\right) .
\end{equation} 
Consequently, the discounted real world value of the annuity at the time
$t \in \{\textrm{Jan }1932, \textrm{Feb }$ $1932, \dots, \textrm{Dec }1971\}$ equals
\begin{equation} 
 \label{val532a}
{\bar{V}}^{\textrm{RW}}_t=\sum_{T \in G}\frac{S^*_t}{S^0_t} E_t\Big(\frac{S^0_T}{S^*_T}\Big)=
\sum_{T\in G}\left( 1-\exp\left\{-\frac{2\,\eta\,{\bar{S}}^*_t}
 {\alpha\,(\exp\{\eta\,T\} - \exp\{\eta\,t\})} \right\}\right) 
\end{equation} 
for the set of payment dates $G=\{\textrm{Jan }1972,\textrm{Jan }1973,\dots,\textrm{Jan }2016\}$. For the previously fitted parameters of the MMM, Figure~\ref{fig:val56} shows the discounted price of the fair annuity (denominated in units of the savings account) according to formula \eqref{val532a},
as a function of the purchasing time $t$. One notes that the time of purchase plays a significant role. Over the years the discounted fair annuity becomes more expensive. The value of the fair annuity remains always below that of the corresponding risk neutral value. It is ranging from about 10\% of the risk neutral value in January 1932 to about 89\% of the risk neutral value in December 1971. This means, someone who starts at the beginning of her or his working life to prepare for retirement can enjoy benefits about eight times greater than those delivered from a later start which is close to retirement. Note that the typical compounding effect in a savings account does not matter in this example, because the value of the annuity and its payments are denominated in units of the savings account. It is the exploitation of the strict supermartingale property of the benchmarked savings account that creates the remarkable effect. The payoff stream of the fair annuity needs to be hedged, generating only a small benchmarked profit and loss or hedge error of similar size as demonstrated in the previous section, where we considered a single fair zero coupon bond. Now we have a portfolio of bonds that pays units of the savings account at their maturities. To demonstrate that the above effect holds independently from the period entered, we repeat with the same parameters for the MMM and similar contracts the calculations for all possible start dates within the period from January 1871 until January 1932. We show in Table~\ref{Tab:3} the mean percentage saving by using the proposed benchmark methodology and the standard deviation for this estimate.

\begin{table}
\begin{tabular}{|c|c|c|c|}
\hline
 & Discounted Risk  & Discounted Real  & Percentage \\
 & Neutral Value & World Value & Saving\\
\hline
Mean & 45&	13.672484&	69.616702\\
Standard Deviation & 0&	5.540818&	12.312929\\
\hline
\end{tabular}
\caption{Comparison of discounted values of deferred annuities paying one dollar per annum for forty five years, commencing in 40 years' time, over all monthly start dates in the period January 1871 to January 1932.}
\label{Tab:3}
\end{table}

\subsection{Long Term Mortality and Equity-Linked Annuities with Mortality and Cash-Linked Guarantees}
Consider now a stylized example that aims to illustrate valuation and hedging under the benchmark approach
in the context of annuities that offer optional guarantees. We may assume also here that interest rates are stochastic because this will not affect our valuation of the product. Consider an annuity that pays to a policy holder at times $T\in G$ until maturity the greater value of $MI_T\times \exp(\eta(T-t_0))$ units of the savings account and $MI_T$ units of the S$\&$P500 total return index account per year at the beginning of each year while the policy holder is alive. Here we assume that the potential savings account or the total return index account payments commence with \$1 at the date $t_0$ of purchasing the annuity. As in the previous example, we assume that there is an asset management fee payable as annuity whose payoff at time $T$ is the same as for a living policy holder.  This type of payoff is likely to account well in the long run for the effect of inflation and equity returns.

The real world value at time $t$ of a portfolio of payments at the future time $T$ is given by
\begin{equation}
V^{\textrm{RW}}_{t,T} = S_t^* E_t\left( \frac{1}{S_T^*}\max \Big( \frac{S^*_T}{S^*_{t_0}}, \frac{S^0_T\exp(\eta(T-t_0))}{S^0_{t_0}}   \Big)   \right)
\end{equation}
which rearranges as
\begin{align}
\label{Eqn:ValueEquityLinkedAnnuity}
V^{\textrm{RW}}_{t,T} &= \frac{S^0_t}{S^0_{t_0}} E_t\left( 
\max \Big( \frac{{\bar{S}}^*_{t}}{{\bar{S}}^*_{t_0}} , \frac{{\bar{S}}^*_{t}\exp(\eta(T-t_0))}{{\bar{S}}^*_T} \Big)   
\right) \\ \notag
&= \frac{{{S}}^*_{t}}{{{S}}^*_{t_0}} 
+ \frac{S^0_t}{S^0_{t_0}} \exp(\eta(T-t_0)) E_t
\left( 
\Big(  \frac{{\bar{S}}^*_{t}}{{\bar{S}}^*_T}  - \frac{{\bar{S}}^*_{t}}{{\bar{S}}^*_{t_0}\exp(\eta(T-t_0))} \Big)^{+}   
\right) .
\end{align}
This rearrangment shows that today's value is the sum of an equity-linked component and an equity index put option component, where the discounted index value is non-central chi-squared distributed. As in the previous example, neither the interest rate nor the mortality rate plays a role in the valuation formula.

We use the following lemma to compute the price of a guarantee.
\begin{lemma} 
\label{Lem:ncx2}
Let $U$ be a non-central chi-squared random variable with four degrees of freedom and non-centrality parameter $\lambda>0$. Then the following expectations hold:
\begin{align}
E\left( \frac{\lambda}{U} \right) &= 1-\exp (-\lambda /2)\\ 
E\left( \frac{\lambda}{U} 1_{U< x}\right) &= \chi^2_{0,\lambda }(x) - \exp (-\lambda /2)\\ 
E\left( \frac{\lambda}{U} 1_{U\ge x}\right) &= 1-\chi^2_{0,\lambda }(x) ,
\end{align}
where $\chi^2_{\nu,\lambda}(x)$ denotes the cumulative distribution function of a non-central chi-squared random variable having $\nu$ degrees of freedom and non-centrality parameter $\lambda$.
\end{lemma}

The proof of this result is given in Appendix~\ref{App:B}.

Furthermore, we mention the following result which we prove in Appendix~\ref{App:C}.

\begin{corollary}
\label{Cor:expn}
For a discounted num\'eraire portfolio ${\bar{S}}^*_{t}$ obeying the SDE~\eqref{lm2.8'} we have the expectations
\begin{align}
E_t\left( \frac{{\bar{S}}^*_{t}}{{\bar{S}}^*_T} \right) &= 1-\exp (-\lambda /2)\\ 
E_t\left( \frac{{\bar{S}}^*_{t}}{{\bar{S}}^*_T} 1_{{\bar{S}}^*_T< x}\right) &= \chi^2_{0,\lambda }\left(\frac{x}{\varphi_T - \varphi_t}\right) - \exp (-\lambda /2)\\ 
E_t\left( \frac{{\bar{S}}^*_{t}}{{\bar{S}}^*_T} 1_{{\bar{S}}^*_T\ge x}\right) &= 1-\chi^2_{0,\lambda }\left(\frac{x}{\varphi_T - \varphi_t}\right) ,
\end{align}
where $\chi^2_{\nu,\lambda}(x)$ denotes the cumulative distribution function of a non-central chi-squared random variable having $\nu$ degrees of freedom and non-centrality parameter $\lambda$ given by
\begin{equation}
\lambda = \frac{{\bar{S}}^*_{t}}{\varphi_T - \varphi_t}
\end{equation}
and $\varphi_t = \frac{1}{4\eta}\alpha(\exp(\eta t)-1)$.
\end{corollary}

This leads us to the following result, which we prove in Appendix~\ref{App:D}.

\begin{theorem}
\label{Thm:D}
For a discounted num\'eraire portfolio ${\bar{S}}^*_{t}$, obeying the SDE~\eqref{lm2.8'}, we have
\begin{align}
&E_t \left( 
\Big(  \frac{{\bar{S}}^*_{t}}{{\bar{S}}^*_T}  - \frac{{\bar{S}}^*_{t}}{{\bar{S}}^*_{t_0}\exp(\eta(T-t_0))} \Big)^{+}   
\right) \\ \notag
&= \chi^2_{0,\lambda }\left(\kappa \right) - \exp (-\lambda /2)
- \frac{{\bar{S}}^*_{t}}{{\bar{S}}^*_{t_0}\exp(\eta(T-t_0))} \chi^2_{4,\lambda }\left(\kappa \right) ,
\end{align}
where $\chi^2_{\nu,\lambda}(x)$ denotes the cumulative distribution function of a non-central chi-squared random variable having $\nu$ degrees of freedom and non-centrality parameter $\lambda$ given by
\begin{equation}
\lambda = \frac{{\bar{S}}^*_{t}}{\varphi_T - \varphi_t}
\end{equation}
and
\begin{align}
\kappa &= \frac{{\bar{S}}^*_{t_0}\exp(\eta(T-t_0))}{\varphi_T - \varphi_t} \\
\varphi_t &= \frac{1}{4\eta}\alpha(\exp(\eta t)-1).
\end{align}
\end{theorem}

Therefore, we can calculate $V^{\textrm{RW}}_{t,T}$ in \eqref{Eqn:ValueEquityLinkedAnnuity} as
\begin{align}
V^{\textrm{RW}}_{t,T} =&   \frac{{{S}}^*_{t}}{{{S}}^*_{t_0}} \left( 1 - \chi^2_{4,\lambda(t,T)} (\kappa(t,T)) \right) \\ \notag
&+ \frac{S^0_t\exp (\eta (T-t_0))}{S^0_{t_0}} \left( \chi^2_{0,\lambda(t,T)} (\kappa(t,T)) -\exp(-\lambda(t,T) /2)   \right) ,
\end{align}
where
\begin{align}
\lambda (t,T) &= \frac{{\bar{S}}^*_{t} }{\varphi_T - \varphi_t} \\ \notag
\kappa(t,T) &= \frac{{\bar{S}}^*_{t_0}\exp(\eta (T-t_0)) }{ \varphi_T - \varphi_t} \\ \notag
\varphi_t &= \frac{1}{4\eta}\alpha(\exp(\eta t)-1) .
\end{align}
Summing over all payment dates $T\in G$ gives the value of the annuity
\begin{align}
V^{\textrm{RW}}_{t,T} =&   \frac{{{S}}^*_{t}}{{{S}}^*_{t_0}} \sum_{T\in G}\left( 1 - \chi^2_{4,\lambda(t,T)} (\kappa(t,T)) \right) \\ \notag
&+ \frac{S^0_t}{S^0_{t_0}} \sum_{T\in G}\exp (\eta (T-t_0))\left( \chi^2_{0,\lambda(t,T)} (\kappa(t,T)) -\exp(-\lambda(t,T) /2)   \right).
\end{align}

\begin{figure}
  
  \centering
    \includegraphics[width=\textwidth]{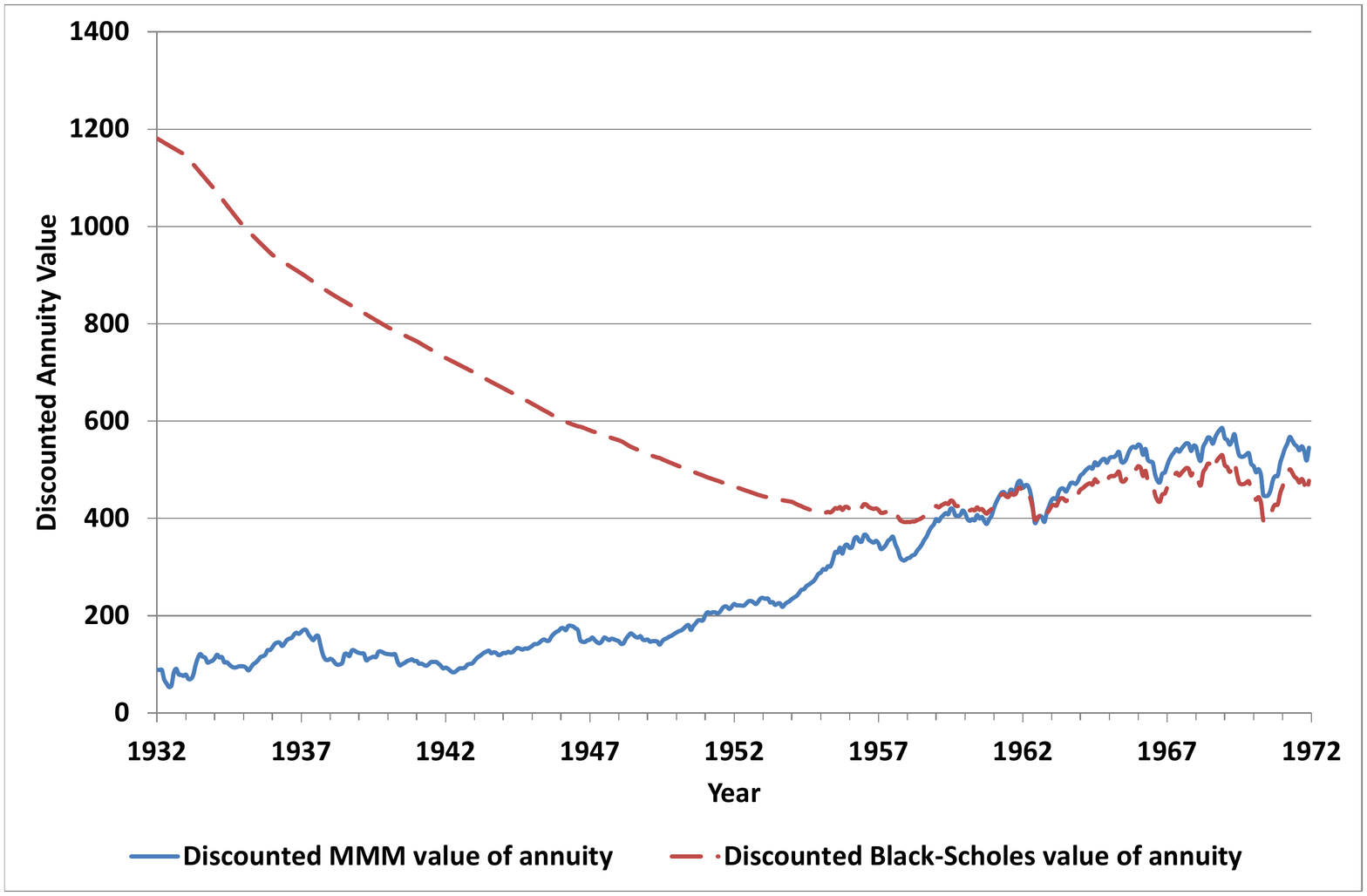}
	\caption{Savings account discounted values under the MMM and the Black-Scholes model of the mortality and equity-linked annuity with mortality and cash-linked guarantee during the period from 1932 until 1971, which provides annual payments from 1972 until 2016.}    
	\label{fig:val57}
\end{figure}

For the previously fitted parameters of the MMM, Figure~\ref{fig:val57} shows the discounted price of the fair annuity (denominated in units of the savings account) according to the formula
\begin{equation} 
 \label{val535b}
 \bar{V}^{\textrm{RW}}_t=\frac{V^{\textrm{RW}}_t}{S^0_t}
\end{equation} 
in dependence on the purchasing time $t$.

Also, for the sake of comparison, the discounted price of the annuity is shown in Figure~\ref{fig:val57} under the assumption of geometric Brownian motion of the discounted num\'eraire portfolio $\bar{S}^*$, that is, a Black-Scholes dynamics with SDE
\begin{equation}
\label{ass:GBM}
d\bar{S}^*_t = \theta^2 \bar{S}^*_t dt + \theta \bar{S}^*_t dW_t,
\end{equation} 
where $\theta $ is estimated using maximum likelihood estimation (MLE) as $\BStheta$ with standard error $\BSthetaSE$ and log-likelihood $\BSloglkhd$ (see, for example, \cite{Fergusson17b}). We can use \eqref{dt4'.b7} to price the annuity because under \eqref{ass:GBM} the Radon-Nikodym derivative $\Lambda_t = \frac{S^0_t\,S^*_{t_0}}{S^*_t\,S^0_{t_0}}$ is a martingale. We have the following theorem.

\begin{theorem}
\label{Thm:E}
For a discounted num\'eraire portfolio ${\bar{S}}^*_{t}$ obeying the SDE~\eqref{ass:GBM} we have
\begin{align}
&E_t \left( 
\Big(  \frac{{\bar{S}}^*_{t}}{{\bar{S}}^*_T}  - \frac{{\bar{S}}^*_{t}}{{\bar{S}}^*_{t_0}\exp(\eta(T-t_0))} \Big)^{+}   
\right) \\ \notag
&=  N(d_1) - \frac{{\bar{S}}^*_{t}}{{\bar{S}}^*_{t_0}\exp(\eta (T-t_0))} N(d_2) ,
\end{align}
where $N(x)$ denotes the cumulative distribution function of a standard normal random variable and $d_1$ and $d_2$ are given by
\begin{align}
d_1 &= \frac{\log\left( \frac{{\bar{S}}^*_{t_0}\exp(\eta (T-t_0))}{{\bar{S}}^*_{t}}\right) + \frac{1}{2}\theta^2(T-t)}{\theta\sqrt{T-t}} \\
d_2 &= d_1 -\theta\sqrt{T-t}.
\end{align}
\end{theorem}

See Appendix~\ref{App:E} for a proof of this theorem.

Thus the value of the annuity under Black-Scholes dynamics has the formula
\begin{equation}
{V}^{\textrm{RN}}_t = \sum_{T\in G} \left( \frac{{{S}}^*_{t}}{{{S}}^*_{t_0}} (1-N(d_2(t,T)))
+ \frac{S^0_t}{S^0_{t_0}} \exp(\eta(T-t_0)) N(d_1(t,T))
\right) 
\end{equation}
and the discounted value of the annuity has the formula
\begin{equation}
\bar{V}^{\textrm{RN}}_t = \frac{1}{S^0_{t_0}} \sum_{T\in G} \left( \frac{{\bar{S}}^*_{t}}{{\bar{S}}^*_{t_0}} (1-N(d_2(t,T)))
+  \exp(\eta(T-t_0)) N(d_1(t,T))
\right)  
\end{equation}
where 
\begin{align}
d_1(t,T) &= (\log (\bar{S}^*_{t_0}\exp(\eta(T-t_0))/\bar{S}^*_{t}) + \frac{1}{2}\theta^2 (T-t))/\sqrt{\theta^2 (T-t)}\\ \notag
d_2(t,T) &= d_1(t,T) - \sqrt{\theta^2 (T-t)} .
\end{align}
Here $N(x)$ is the cumulative distribution function of the standard normal distribution.
The discounted real world price of the annuity under the MMM has the initial value 88.853 which makes it significantly less expensive than the initial value 1181.076 for the discounted price of the annuity when assuming the Black-Scholes model for the index dynamics.

The above examples are deliberately designed to illustrate the cost effectiveness of real world valuation of long term contracts under the MMM compared to valuation under the classical market model and paradigm. It is obvious that the introduction of mortality risk and more refined models for the index dynamics would not materially change the principal message provided by these preceding examples: There are less expensive ways to transfer wealth into payoff streams than classical modeling and valuation approaches can provide.

 \section{Conclusion}
 
   \label{sec:val10}
The paper proposes to move away from classical risk neutral valuation towards
a more general real world valuation methodology under the benchmark approach.
The resulting production methodology does not assume the existence of an equivalent risk neutral
probability measure and offers, therefore, a much wider modeling world.
As a consequence, the better long term performance of the
equity market compared to that of the fixed income market can be systematically exploited to produce less expensively pension and insurance products.
Real world valuation allows one to
generate hedge portfolios with
prices for long term contracts that can be significantly lower than those obtained under classical pricing paradigms. The proposed real world valuation methodology uses the best performing portfolio, the num\'eraire portfolio, as num\'eraire and the real world probability measure as pricing measure when taking expectations. Real world valuation identifies
the minimal possible value for a replicable contingent claim. 
Real
world pricing generalizes classical risk neutral pricing and also
actuarial pricing.

\appendix
\renewcommand \thesection {\Alph{section}}

\section{Maximum Likelihood Estimation of MMM Parameters}
\label{App:A}
Given a series of observations of the discounted index
\begin{equation}
{\bar{S}}^*_{t_0} , {\bar{S}}^*_{t_1} , \ldots , {\bar{S}}^*_{t_n}
\end{equation}
at the times $t_0<t_1<\ldots <t_n$
we seek the values of the parameters $\alpha$ and $\eta$ of the SDE
 \begin{equation} 
 d\, {\bar{S}}^*_t =\alpha_t\,dt + \sqrt{{\bar{S}}^*_t\,\alpha_t}
 \,dW_t,
 \end{equation} 
which maximize the likelihood of the occurrence of the observations under the hypothesis that the stylized version of the
MMM holds.
Here we have $t\ge 0$, ${\bar{S}}^*_0 >0$, $W=\{W_t, t\ge 0\}$ being a Wiener process and $\alpha_t$
modeled by the exponential function
\begin{equation} 
\alpha_t= {\alpha}\,\exp\{\eta\,t\}.
\end{equation} 
The transition density of the discounted num\'eraire portfolio ${\bar{S}}^*$ is given in~\eqref{Eqn:transitiondensityMMMdiscountedGOP}.
Using this transition density function the logarithm of the likelihood function, which we seek to maximize, is found to be
\begin{align}
&\ell (\alpha, \eta) \\ \notag
&= \sum_{i=1}^n \log \bigg( \frac{1}{2(\varphi_{t_i} - \varphi_{t_{i-1}} )} \sqrt{\frac{{\bar{S}}^*_{t_i}}{{\bar{S}}^*_{t_{i-1}}}} \exp \left( -\frac{{\bar{S}}^*_{t_{i-1}} + {\bar{S}}^*_{t_i}}{2(\varphi_{t_i} - \varphi_{t_{i-1}} )} \right) I_1 \left( \frac{\sqrt{{\bar{S}}^*_{t_{i-1}} {\bar{S}}^*_{t_i}}}{\varphi_{t_i} - \varphi_{t_{i-1}}} \right) \bigg) \\ \notag
&= \sum_{i=1}^n \bigg\{ \log\bigg(\frac{1}{2(\varphi_{t_i} - \varphi_{t_{i-1}} )}\bigg) +\frac{1}{2}\log \bigg(\frac{{\bar{S}}^*_{t_i}}{{\bar{S}}^*_{t_{i-1}}}\bigg) + \left( -\frac{{\bar{S}}^*_{t_{i-1}} + {\bar{S}}^*_{t_i}}{2(\varphi_{t_i} - \varphi_{t_{i-1}} )} \right) \\ \notag
&+ \log \bigg( I_1 \left( \frac{\sqrt{{\bar{S}}^*_{t_{i-1}} {\bar{S}}^*_{t_i}}}{\varphi_{t_i} - \varphi_{t_{i-1}}} \right) \bigg) \bigg\}.
\end{align}
Initial estimates of $\alpha $ and $\eta$ can be found by equating the empirically calculated quadratic variation of $\sqrt{{\bar{S}}^*}$, that is
\begin{equation}
\langle \sqrt{{\bar{S}}^*} \rangle_{t_j} = \sum_{i=1}^j (\sqrt{{\bar{S}}^*_{t_i}} - \sqrt{{\bar{S}}^*_{t_{i-1}}})^2,
\end{equation}
to the theoretical quadratic variation of $\sqrt{{\bar{S}}^*}$, that is
\begin{equation}
\langle \sqrt{{\bar{S}}^*} \rangle_{t_j} = \frac{\alpha}{4\eta}(\exp (\eta t_j) - 1),
\end{equation}
at the times $t=t_k$ and $t=t_{2k}$ where $k=\lfloor n/2\rfloor$. The initial estimates are found straightforwardly to be
\begin{align}
\alpha_0 &= \langle \sqrt{{\bar{S}}^*} \rangle_{t_k} \frac{4\eta}{\exp(\eta t_k)-1} \\ \notag
\eta_0 &= \log \frac{\langle \sqrt{{\bar{S}}^*} \rangle_{t_{2k}}/\langle \sqrt{{\bar{S}}^*} \rangle_{t_k} - 1}{t_k-t_0}.
\end{align}
With these initial estimates we calculate the logarithm of the likelihood function at points $\alpha_0 + i \delta \alpha$ and $\eta_0 + j\delta \eta$ for $i,j=-2,-1,0,1,2$, $\delta\alpha = \alpha_0/4$ and $\delta\eta = \eta_0 / 4$ and fit a quadratic form
\begin{equation}
Q(x) = x^T A x - 2b^T x + c ,
\end{equation}
where $x = \left( \begin{array}{c} \alpha \\ \eta \end{array}\right) $ is a 2-by-1 vector, $A$ is a negative definite 2-by-2 matrix, $b$ is a 2-by-1 vector and $c$ is a scalar.
Subsequent estimates $\alpha_1 $ and $\eta_1$ are obtained as the matrix expression $A^{-1} b$ corresponding to the maximum of the quadratic form $Q$. Iteratively applying this estimation method gives the maximum likelihood estimates of the parameters $\alpha$ and $\eta$:
\begin{align}
\alpha &= \MMMalphabar \, (\MMMalphabarSE ), \\ \notag
\eta &= \MMMetaSE\, (\MMMetaSE ) ,
\end{align}
where the standard errors are shown in brackets. The maximum log likelihood is $\ell (\alpha, \eta) = \MMMloglkhd $. The Cram\'er-Rao inequality for the covariance matrix of the parameter estimates is
\begin{equation}
VAR((\alpha,\eta)) \ge -\frac{1}{\nabla^2 \ell(\alpha,\eta)}
\end{equation}
and as the number of observations becomes large the covariance matrix approaches the lower bound, which we use to calculate the standard errors.

\section{Proof of Lemma~\ref{Lem:ncx2}}
\label{App:B}

\begin{refproof*}[of Lemma~\ref{Lem:ncx2}]
If $U$ is a non-central chi-squared random variable having $\nu$ degrees of freedom and non-centrality parameter $\lambda$, then it can be written as a chi-squared random variable $X_N$ having a random number of degrees of freedom $N=\nu+2P$ with $P$ being a Poisson random variable with mean $\lambda /2$, see e.g.~\cite{JohnsonKoBa95}. It follows that
\begin{align}
E\left( \frac{\lambda}{U}1_{U<x}\right) &= E\left( \frac{\lambda}{X_N}1_{X_N<x}\right) \\ \notag
&= E\left( E\left( \frac{\lambda}{X_N}1_{X_N<x}\bigg| N\right) \right) .
\end{align}
Because $X_N $ is conditionally chi-squared distributed we have that 
\begin{equation}
E\left(\frac{1}{X_N} 1_{X_N<x} \bigg|N\right) = \frac{1}{N-2} E(1_{X_{N-2}<x}|N)
\end{equation}
for a conditionally chi-squared random variable $X_{N-2}$ with $N-2$ degrees of freedom.
Therefore
\begin{equation}
E\left( \frac{\lambda}{U}1_{U<x}\right) = E\left( \frac{\lambda }{N-2} E(1_{X_{N-2}<x}|N) \right) 
\end{equation}
and substituting $\nu + 2P$, with $\nu=4$, for $N$ gives
\begin{equation}
\label{Eqn:expn}
E\left( \frac{\lambda}{U}1_{U<x}\right) = \frac{1}{2} E\left( \frac{\lambda }{P+1} E(1_{X_{2+2P}<x}|P) \right) .
\end{equation}
We observe that for any Poisson random variable $P$ with mean $\mu$,
\begin{equation}
E\left( \frac{1}{1+P} f(P)\right) = \frac{1}{\mu} E\left( f(P-1)\right) - \frac{1}{\mu}\exp(-\mu) f(-1)
\end{equation}
and making use of this, \eqref{Eqn:expn} becomes, with $\mu = \lambda/2$ and $f(P) = E(1_{X_{2+2P}<x}|P)$,
\begin{align}
\label{Eqn:expn2}
E\left( \frac{\lambda}{U}1_{U<x}\right) &= \frac{\lambda}{2} \left(\frac{1}{\mu} E\left( f(P-1)\right) - \frac{1}{\mu}\exp(-\mu) f(-1)\right) \\ \notag
&= E\left( E(1_{X_{2P}<x}|P)\right) - \exp(-\lambda /2) \\ \notag
&= \chi^2_{0,\lambda}(x) - \exp(-\lambda /2),
\end{align}
which is the second expectation formula.
Letting $x\to\infty$ in \eqref{Eqn:expn2} gives the first expectation formula.
The third expectation formula follows straightforwardly by subtracting the second formula from the first formula.
\end{refproof*}

\section{Proof of Corollary~\ref{Cor:expn}}
\label{App:C}

\begin{refproof*}[of Corollary~\ref{Cor:expn}]
Letting 
\begin{equation}
\lambda = \frac{{\bar{S}}^*_{t}}{\varphi_T - \varphi_t}
\end{equation}
we know that the distribution of the random variable
\begin{equation}
U = \frac{{\bar{S}}^*_{T}}{\varphi_T - \varphi_t}
\end{equation}
conditional upon ${\bar{S}}^*_{t}$ is a non-central chi-squared distribution with four degrees of freedom and non-centrality parameter $\lambda$. Applying Lemma~\ref{Lem:ncx2} gives the result.
\end{refproof*}

\section{Proof of Theorem~\ref{Thm:D}}
\label{App:D}

\begin{refproof*}[of Theorem~\ref{Thm:D}]
Since 
\begin{equation}
\frac{{\bar{S}}^*_{t}}{{\bar{S}}^*_T}  - \frac{{\bar{S}}^*_{t}}{{\bar{S}}^*_{t_0}\exp(\eta(T-t_0))} > 0
\end{equation}
if and only if
\begin{equation}
{\bar{S}}^*_T <  {\bar{S}}^*_{t_0}\exp(\eta(T-t_0)) 
\end{equation}
we can write
\begin{align}
&E_t \left( 
\Big(  \frac{{\bar{S}}^*_{t}}{{\bar{S}}^*_T}  - \frac{{\bar{S}}^*_{t}}{{\bar{S}}^*_{t_0}\exp(\eta(T-t_0))} \Big)^{+}   
\right) \\ \notag
&= E_t \left( 
  \frac{{\bar{S}}^*_{t}}{{\bar{S}}^*_T} 1_{{\bar{S}}^*_T <  {\bar{S}}^*_{t_0}\exp(\eta(T-t_0))} \right) 
  - E_t \left( \frac{{\bar{S}}^*_{t}}{{\bar{S}}^*_{t_0}\exp(\eta(T-t_0))}1_{{\bar{S}}^*_T <  {\bar{S}}^*_{t_0}\exp(\eta(T-t_0))} \right) .
\end{align}
Applying Corollary~\ref{Cor:expn} and the fact that ${\bar{S}}^*_T$ is non-central chi-squared distributed with four degrees of freedom gives
\begin{align}
&E_t \left( 
\Big(  \frac{{\bar{S}}^*_{t}}{{\bar{S}}^*_T}  - \frac{{\bar{S}}^*_{t}}{{\bar{S}}^*_{t_0}\exp(\eta(T-t_0))} \Big)^{+}   
\right) \\ \notag
  &= \chi^2_{0,\lambda }\left(\kappa\right) - \exp (-\lambda /2) - \frac{{\bar{S}}^*_{t}}{{\bar{S}}^*_{t_0}\exp(\eta(T-t_0))} \chi^2_{4,\lambda }\left(\kappa \right)
\end{align}
where
\begin{equation}
\kappa = \frac{{\bar{S}}^*_{t_0}\exp(\eta(T-t_0))}{\varphi_T - \varphi_t},
\end{equation}
which is the requested result.
\end{refproof*}

\section{Proof of Theorem~\ref{Thm:E}}
\label{App:E}

\begin{refproof*}[of Theorem~\ref{Thm:E}]
The proof is analogous to that of the Black-Scholes European call option formula. We can write $\frac{{\bar{S}}^*_{t}}{{\bar{S}}^*_T}$ as
\begin{equation}
\frac{{\bar{S}}^*_{t}}{{\bar{S}}^*_T} = \exp( -\frac{1}{2}\theta^2(T-t)+\theta\sqrt{T-t}Z)
\end{equation}
for a standard normally distributed random variable $Z$. 
Thus the condition
\begin{equation}
\frac{{\bar{S}}^*_{t}}{{\bar{S}}^*_T}  - \frac{{\bar{S}}^*_{t}}{{\bar{S}}^*_{t_0}\exp(\eta(T-t_0))} > 0
\end{equation}
is equivalent to the condition
\begin{equation}
Z >  \frac{\log\left( \frac{{\bar{S}}^*_{t}\exp(\frac{1}{2}\theta^2(T-t))}{{\bar{S}}^*_{t_0}\exp(\eta (T-t_0))}\right) }{\theta\sqrt{T-t}} = -d_2.
\end{equation}
We can rewrite the expectation as
\begin{align}
&E_t\left(  \exp( -\frac{1}{2}\theta^2(T-t)+\theta\sqrt{T-t}Z) 1_{Z>-d_2} - \frac{{\bar{S}}^*_{t}}{{\bar{S}}^*_{t_0}\exp(\eta(T-t_0))} 1_{Z>-d_2}  \right) \\ \notag
&= \exp( -\frac{1}{2}\theta^2(T-t)) E_t\left(  \exp( \theta\sqrt{T-t}Z) 1_{Z>-d_2}\right) \\ \notag
&- \frac{{\bar{S}}^*_{t}}{{\bar{S}}^*_{t_0}\exp(\eta(T-t_0))} E_t\left( 1_{Z>-d_2} \right) .
\end{align}
We note that $E_t(\exp(\alpha Z)1_{Z>-d_2}) = \exp(\frac{1}{2}\alpha^2)(1-N(-d_2-\alpha))$ so that the expectation becomes
\begin{equation}
(1-N(-d_2-\theta\sqrt{T-t})) - \frac{{\bar{S}}^*_{t}}{{\bar{S}}^*_{t_0}\exp(\eta(T-t_0))} (1-N(-d_2)) ,
\end{equation}
which simplifies to the result.
\end{refproof*}

\bibliographystyle{plainnat}

\bibliography{HedgingCLBsRevised}

\newpage

\end{document}